\documentclass[runningheads,a4paper]{llncs}  

\usepackage{amssymb}  
\usepackage{braket}   
\usepackage{graphicx}
\usepackage[french,english]{babel}   
\usepackage[style=french]{csquotes}  

\newcommand{\ob}{\linebreak[0]}      

\newcommand{\Delt}{\mathrm\Delta}            
\newcommand{\lp}{Landauer's Principle}       
\newcommand{\nc}{non-com\-pu\-ta\-tion\-al}  

\newcommand{\ie}{\emph{i.e.}}						
\newcommand{\Ie}{\emph{I.e.}}
\newcommand{\eg}{\emph{e.g.}}

\newcommand{\etal}{\emph{et al.}}

\newcommand{\etc}{\emph{etc.}}

\usepackage{ragged2e}
\usepackage{fancyhdr}
\fancypagestyle{title}{%
  \setlength{\headheight}{23pt}
  \fancyhf{}
  \fancyhead[C]{\textsc{Extended author's postprint of a paper previously 		%
  		published$^\dagger$ in the Proc.\ of the 10th Conf.\ on Reversible 		%
  		Computation (RC18)}}%
  \fancyfoot[L]{\justify\hangindent=1em $^\dagger$Published version: Michael 	%
  			P. Frank, ``Physical Foundations of Landauer's Principle,'' in 		%
  			Jarkko Kari and Irek Ulidowski, eds., {\emph{Reversible 				%
  			Computation: 10th International Conference, RC 2018, Leicester, 	%
  			UK, September 12--14, 2018}}, Lecture Notes in Computer Science, 	%
  			vol.{\ }11106, pp.{\ }3--33, Springer International Publishing, 	%
  			Cham, 2018. {\doi{10.1007/978-3-319-99498-7\_1}}}%
}%

\begin{document}

\mainmatter

\title{Physical Foundations of Landauer's Principle%
\thanks{This work was supported by the Laboratory Directed Research and 
        Development pro\-gram at Sandia National Laboratories and by the 
        Advanced Simulation and Computing program under the U.S. Department 
        of Energy's National Nuclear Security Administration (NNSA).  Sandia 
        National Laboratories is a multimission laboratory managed and 
        operated by National Technology and Engineering Solutions of Sandia, 
        LLC., a wholly owned subsidiary of Honeywell International, Inc., for 
        NNSA under con\-tract DE-NA0003525.  Approved for public release,
        SAND2019-0892 O.  This paper describes objective technical results and 
        analysis.  Any subjective views or opinions that might be expressed in 
        this paper do not necessarily represent the views of the U.S.{\ }%
        Department of Energy or the United States Government.}%
}

\author{Michael P. Frank}
\authorrunning{M. Frank}

\institute{Center for Computing Research, Sandia National Laboratories,\\
           P.O. Box 5800, Mail Stop 1322, Albuquerque, NM 87185\\
\email{mpfrank@sandia.gov}\\
\url{http://www.cs.sandia.gov/cr-mpfrank}}

\maketitle     

\thispagestyle{title}		


\begin{abstract}
We review the physical foundations of {\lp},	which relates the loss of             
information from a computational process to an increase in thermodynamic entropy.  
Despite the long history of the Principle, its fundamental rationale and proper    
interpretation remain frequently misunderstood.  Contrary to some                  
misinterpretations of the Principle, the mere {\emph{transfer}} of                 
entropy between computational and {\nc} subsystems can occur in a                  
thermodynamically reversible way without increasing total entropy.  However,       
{\lp} is not about general entropy transfers; rather, it more specifically         
concerns the ejection of (all or part of) some {\emph{correlated}} information     
from a controlled, digital form ({\eg}, a computed bit) to an uncontrolled,        
{\nc} form, {\ie}, as part of a thermal environment.  Any uncontrolled thermal     
system will, by definition, continually re-randomize the physical information in   
its thermal state, from our perspective as observers who cannot predict the        
exact dynamical evolution of the microstates of such environments.  Thus, any      
correlations involving information that is ejected into and subsequently           
thermalized by the environment will be lost from our perspective, resulting        
directly in an irreversible increase in total entropy.  Avoiding the      			
ejection and thermalization of correlated computational information motivates      
the reversible computing paradigm, although the requirements for computations      
to be thermodynamically reversible are less restrictive than frequently            
described, particularly in the case of stochastic computational operations.        
There are interesting possibilities for the design of computational processes      
that utilize stochastic, many-to-one computational operations while nevertheless   
avoiding net entropy increase that remain to be fully explored.                    


\end{abstract}


\section{Introduction}
\label{sec:intro}

A core motivation for the field of reversible computation is {\lp} 
{\cite{Landauer-61}}, which tells us that each bit's worth of information that            
is lost from a computational process results in a (permanent) increase in 
thermodynamic entropy of $\Delt S \geq k\ln 2$, where $k=k_\mathrm{B}$ is 
Boltzmann's constant,\footnote{Boltzmann's constant $k_\mathrm{B}\approx1.38
        \times 10^{-23}\ \mathrm{J}/\mathrm{K}$, in traditional units.  This 
        constant was actually introduced by Planck in {\cite{Planck-1901}}.               
        We discuss this history further in {\S\ref{sec:ent-hist}}.} 
with the corresponding dissipation of $\Delt E\geq kT\ln 2$ energy to heat, 
where $T$ is the temperature of the heat sink.  By avoiding information loss, 
reversible computation bypasses this limit on the energy efficiency of 
computing, opening the door to a future of potentially unlimited long-term 
improvements in computational efficiency.\footnote{The mathematical fact, not 
        initially fully understood by Landauer, that reversible computational 
        processes can indeed avoid information loss was rigorously demonstrated 
        by Bennett {\cite{Bennett-73}}, using methods anticipated by Lecerf               
        {\cite{Lecerf-63}}.}                                                              

The correctness of {\lp} has recently been empirically validated 
{\cite{Berut-etal-12,Orlov-etal-12,Jun-etal-14,Yan-etal-18}}, but the results of          
these experiments are unsurprising, given that the validity of {\lp} can be 
shown to follow as a rigorous consequence of basic facts of fundamental physics 
that have been known for over a century, ever since pioneering work by such 
luminaries as Boltzmann and Planck revealed the fundamentally statistical nature 
of entropy, summarized in the equation $S = k\log W$ that is emblazoned on 
Boltzmann's tombstone.\footnote{In this equation, $W$ counts the number of 
        distinct microstates consistent with a given macroscopic description of 
        a system.}  
As we will show in some detail, {\lp} follows directly and rigorously from the 
modern statistical-mechanical understanding of thermodynamics (which elaborates 
upon Boltzmann's view), augmented only by a few mathematical concepts from 
information theory, along with the most basic understanding of what is meant by 
a {\emph{computational process}}.  

However, despite this underlying simplicity, certain subtleties regarding the 
proper interpretation of the Principle remain frequently misunderstood; I have 
discussed some of these in earlier papers 
{\cite{Frank-05,Frank-DeBen-16,Frank-RC17,Frank-RC17-preprint,Frank-GRC18}}, and          
will elaborate upon another one in the present paper.  Issues mentioned in these 
works include:


\begin{enumerate}

\item \textbf{\emph{Treatment of stochastic operations.}} It has long been 
understood that {\emph{stochastic}} or randomizing computational operations can 
transfer entropy from a thermodynamic environment to a digital form, reversing 
the usual process considered in discussions of Landauer, in which computational 
entropy is pushed from a digital form out to a thermal environment.  It follows 
from this observation that the act of merely {\emph{transferring}} isolated bits 
of entropy between computational and thermal forms can actually be a 
thermodynamically (albeit not logically) {\emph{reversible}} process.  As an 
illustration of this, I pointed out in 2005 {\cite{Frank-05}} that a stochastic           
computational process that simply re-randomizes an already-random digital bit 
does not necessarily increase thermodynamic entropy, even though this process 
would not be considered logically reversible (injective) in a traditional 
treatment.  Thus, the usual arguments for {\lp} and reversible computing that do 
not address this case are overly simplistic; later, we will discuss how to 
generalize and repair them.

\item \textbf{\emph{Transformations of complex states.}} The fundamental 
physical arguments behind {\lp} are not constrained to dealing only with bits 
(binary digits or two-state systems) per se; they apply equally well to systems 
with any number of states.  In particular, one can even apply them to 
spatially-extended physical systems with very large numbers of states, so that, 
for example, it is possible in principle to adiabatically transform a system 
representing the state of a complex Boolean logic circuit directly from ``old 
state'' to ``new state'' in a single step without incurring any Landauer losses 
related to the number of Boolean logic operations implemented by the circuit.  
An abstract model illustrating this capability in the context of classical, 
chaotic dynamical systems was described in 2016 
{\cite{Frank-DeBen-16,Frank-Chaos}}.                                                      
An example of an adiabatic physical mechanism that can transform states of 
extended logic networks all at once can be found in the Quantum-dot Cellular 
Automata (QDCA or QCA) approach pioneered by Lent {\etal} (see 
{\cite{Lent-etal-94}} and subsequent papers by that group).  However, an 					
analogous approach can also be carried out even in more conventional CMOS 
technology, by encoding complex logic functions as large series/parallel 
switching circuits that are transformed adiabatically in a single (albeit very 
slow) step.

\item \textbf{\emph{Role of conditional reversibility.}}  A third important 
clarification of {\lp} can be found when considering the role of 
{\emph{conditional reversibility}}, which I explained in 
{\cite{Frank-05,Frank-RC17,Frank-RC17-preprint,Frank-GRC18}}, but which was               
already implicitly leveraged by all of the early implementation concepts for 
reversible computation 
{\cite{Likharev-77,Fredkin-Toffoli-82,Drexler-92,Younis-Knight-93}}.                      
The key point is that states that are prevented from arising by design within a 
given computer architecture (construed generally) have zero probability of 
occurring, and therefore make zero contribution to the entropy that is required 
to be ejected from the computational state by {\lp}.  Therefore, it is a 
sufficient logical-level condition for avoiding Landauer's limit if {\emph{only}} 
the set of computational states {\emph{that are actually allowed to occur}} in 
the context of a given design are mapped one-to-one onto new states.  \Ie, 
the machine can be designed in such a way that it would map the other, 
{\emph{forbidden}} states many-to-one without there being any actual 
thermodynamic impact from this, given that those states will never actually 
occur.  This issue was already discussed extensively in {\cite{Frank-RC17}} 				
(and see {\cite{Frank-RC17-preprint}} for proofs of the theorems), so we will 				
not discuss it in great detail in the present paper.			

\item \textbf{\emph{Importance of correlations.}}  At first, it might seem
that the thermodynamic reversibility of certain logically-irreversible, 
stochastic transformations as discussed in point 1 above contradicts {\lp}.  But 
this apparent contradiction is resolved when one realizes that the proper 
subject of {\lp} is {\emph{not}} in fact the ejection of isolated, purely 
{\emph{random}} bits of digital information from a computer.  Such bits are
{\emph{already}} entropy, and merely moving those bits from a stable digital 
form to a rapidly-changing thermal form does not necessarily increase total 
entropy, as we will illustrate with some basic examples.  Rather, what {\lp} 
really concerns is the ejection of {\emph{correlated}} bits from the 
computational state, since a thermal environment cannot be expected to preserve 
those correlations in any way that is accessible to human modeling.  So really, 
it is the {\emph{loss of prior correlations}} that is the ultimate basis for the 
consideration of information loss and entropy increase in {\lp}.  I addressed 
this issue briefly in previous presentations {\cite{Frank-SFI,Frank-Stanford}};           
in this paper, I elaborate on it in more detail.		
\end{enumerate}

The rest of this paper is organized as follows.  Section~\ref{sec:defs} reviews 
some basic mathematical concepts of entropy, information, and computation.  
Section~\ref{sec:infophys} discusses the connection of these concepts with 
physics in detail, and gives examples of physical systems that illustrate 
the fundamental appropriateness of these abstract concepts for modeling the 
practical physical circumstances that we use them to describe. This discussion 
lays bare the fundamental unity between information theory and physical theory, 
in showing that information-theoretic entropy and thermodynamic entropy really 
are {\emph{the exact same concept}} as each other; they are, in fact, the exact 
same epistemological/physical quantity, merely applied at different levels that 
are nonetheless fundamentally interconnected.  We then use this understanding of 
basic physics to prove {\lp}, and discuss its implications for the energy 
efficiency of future reversible and irreversible computing technologies.  
Section~\ref{sec:experiments} briefly reviews some of the existing laboratory 
studies that have validated {\lp} empirically.  Section~\ref{sec:conclusion} 
concludes with some suggestions for future work.


\section{Definitions of Basic Concepts}
\label{sec:defs}

In this section, we begin by reviewing the mathematical definitions of some 
basic concepts from statistics, information theory, and computation that are 
useful for understanding the thermodynamics of computation in general, and 
{\lp} in particular.  Later (in sec.~3), we'll discuss in detail why these 
mathematical concepts are appropriate not just in abstract conceptual scenarios, 
but also for describing real physical circumstances, and give some examples.


\subsection{Some Basic Statistical Concepts}
\label{sec:stat}

First, let us define a few basic statistical concepts that are adequate for our 
purposes.  In the following treatment, we will take a manifestly epistemological 
perspective, since, as we will see, such a perspective is inevitably quite 
central and fundamental to what not only statistics, but also physical modeling 
in general is all about---since, any physical model inevitably concerns what is 
known, or could be known, about a physical system; and this remains the case 
whether we are talking about the actual knowledge of a real observer (\eg, a 
human experimentalist), or about what could in principle be known by a 
hypothetical omniscient modeler, or by any other real or imagined reasoner (\eg, 
an engineered artifact, considered as an observer).  All sciences are concerned 
with the knowable truths in their domain of applicability, and physics, in 
particular, is ultimately simply the study of what is knowable about the 
bottom-most foundations of this physical world that we live in.  Thus a proper 
mathematical account of epistemology is an essential conceptual foundation for 
any science, and the study of the thermodynamics of computation is no exception.

In the following, we begin with the concept of a discrete variable---where 
``variable'' here is meant in the sense of a {\emph{random variable}} in 
statistics, although we avoid that particular terminology, since it begs the 
question of defining randomness, which is somewhat tangential to our purposes.  
Then we go on to motivate and define some basic concepts of 
{\emph{improbability}}, {\emph{surprise}}, and {\emph{probability}}, along with a 
concept that we will think of as the ``psychological weight'' or 
{\emph{heaviness}} of a possible outcome, together with the {\emph{expected value}} 
of a function, and finally {\emph{entropy}} (a concept that falls out naturally 
from the foregoing ones).  This means of deriving the entropy concept provides 
certain conceptual elements that we will find useful in later discussions.  (But, 
we should emphasize that one does not have to take seriously any particular 
theory of psychology to find these concepts useful---they are simply technical 
definitions.)


\subsubsection{Discrete variables.}
To begin, a {\emph{discrete variable}} $V$ is associated with some countable 
set $\mathbf{V} = \{v_i\}$ of mutually exclusive {\emph{values}} $v_1,\, v_2,\,
\ldots \in \mathbf{V}$ that the variable can take on.  For our purposes, 
typically we will work with value sets $\mathbf{V}$ that are finite.  Our 
subject matter, in statistics and information theory, is the quantitative 
analysis of what is {\emph{known}} about the value of some variable(s).  As
usual, the {\emph{knower}}, here, could in general be any real or hypothetical 
reasoner.


\subsubsection{Improbability and probability.}
Suppose all that is initially known regarding a given discrete variable $V$ is 
the cardinality (number of elements) $n=|\mathbf{V}|$ of its set $\mathbf{V}$ of 
possible values.  Assume $n$ is finite; we write $\mathbf{V} = \{v_1,\, v_2,\, 
\ldots,\, v_n\}$.  Now suppose we somehow subsequently learn that the variable 
has a particular value, $V=v_i$, for some $i\in\{1,\, 2,\, \ldots,\, n\}$.  We 
can say that this particular {\emph{outcome}} or {\emph{event}} (of the learned 
value turning out to be $v_i$) has, {\emph{a priori}}, from the learner's 
perspective, a baseline {\emph{improbability}} $m = m_i = m(v_i)$ given by $m_i 
= \overline{m} = n$, since the more different values $v_i$ there are, the more 
unlikely or {\emph{improbable}} each individual value would seem to be, 
proportionally---not knowing anything else about the situation.  We can then 
define the baseline {\emph{probability}} $p = p_i = p(v_i) = \overline{p}$ of 
each value as the reciprocal of its improbability $m = \overline{m}$, {\ie}, 
$p_i = 1/m_i = 1/n$; note that this derivation yields the usual property that 
the probabilities of all the values $\{v_i\}$ sum to 1, {\ie}, $\sum_{i=1}^n p_i 
= 1$.  


\subsubsection{Surprise, or increase of knowledge.}
We can then quantify the {\emph{amount of increase}} in our knowledge resulting 
from this learning event as our {\emph{surprise}}, or the {\emph{surprisingness}} 
of the event, defined as $s = s_i = s(v_i)=\log m_i=\log 1/p_i=-\log p_i$ 
(dimensioned in general logarithmic units; see~\cite{Frank-arxiv-05}), with the           
motivation for this definition being that ``surprise'' should combine additively 
whenever the number of possible values combines multiplicatively.  

For example, when rolling a 6-sided die, each outcome has an improbability $m$ 
of 6, and the surprise for each case (rolling a 1, say) is then $s = \log 6.$  
If I roll the die twice, there are $6^2 = 36$ possible sequences of outcomes, 
but each of these sequences (say rolling two 1's) is, intuitively, only twice as 
surprising ($\log 36 = 2 \log 6$) as each individual result was in the 1-die 
case.  In any event, regardless of whether the behavior of this definition 
matches your personal intuition about how surprisingness ought to work, 
psychologically, let this be our technical definition of ``surprise.''


\subsubsection{Nonuniform probability distributions.}
If we happen to have more knowledge about the value of the variable than just 
its cardinality, this can be modeled by assigning different probabilities $p_i$ 
(and corresponding improbability and surprise) to different values $v_i$, 
subject to the constraint that the probabilities of all the values are still 
non-negative, and still sum to 1.\footnote{The rule that probabilities must 
        always sum to 1 can be derived by considering the implications, under 
        our definitions, of breaking down all possible events (regardless of 
        their probability) into a set of equally-likely micro-alternatives; only 
        the probability distributions that sum to 1 turn out to be 
        epistemologically self-consistent in that scenario, but we will not 
        detail that argument here.}  
We call the entire function $P:\mathbf{V}\rightarrow[0,1]$ with $P(v_i) = p_i$ 
(over all $i=1,\, 2,\, \ldots,\, n$) a {\emph{probability distribution over $V$,}} 
and write it as $P(V)$.  In this case, the improbabilities $m_i=1/p_i$ and 
surprisingnesses $s_i=\log m_i$ would be adjusted accordingly.

The semantic interpretation of probabilities in this general case can be 
inherited from the ``surprise'' concept; for example, if a particular value $v_i$ 
has probability 1/2, this would mean that its surprise is log 2, and this says 
that our state of knowledge about the variable is such that, if we were to learn 
that it had the value $v_i$, we would be equally surprised as we would have been 
if initially we only knew that it had exactly two possible values, and then we 
suddenly learned that its actual value was one of those.  Thus, this way of 
motivating the concept of probability rests on an intuitive psychological 
interpretation.  


\subsubsection{Heaviness, or ``psychological weight.''}
\begin{figure}[!tb]
        \centering\includegraphics[height=1.6in]{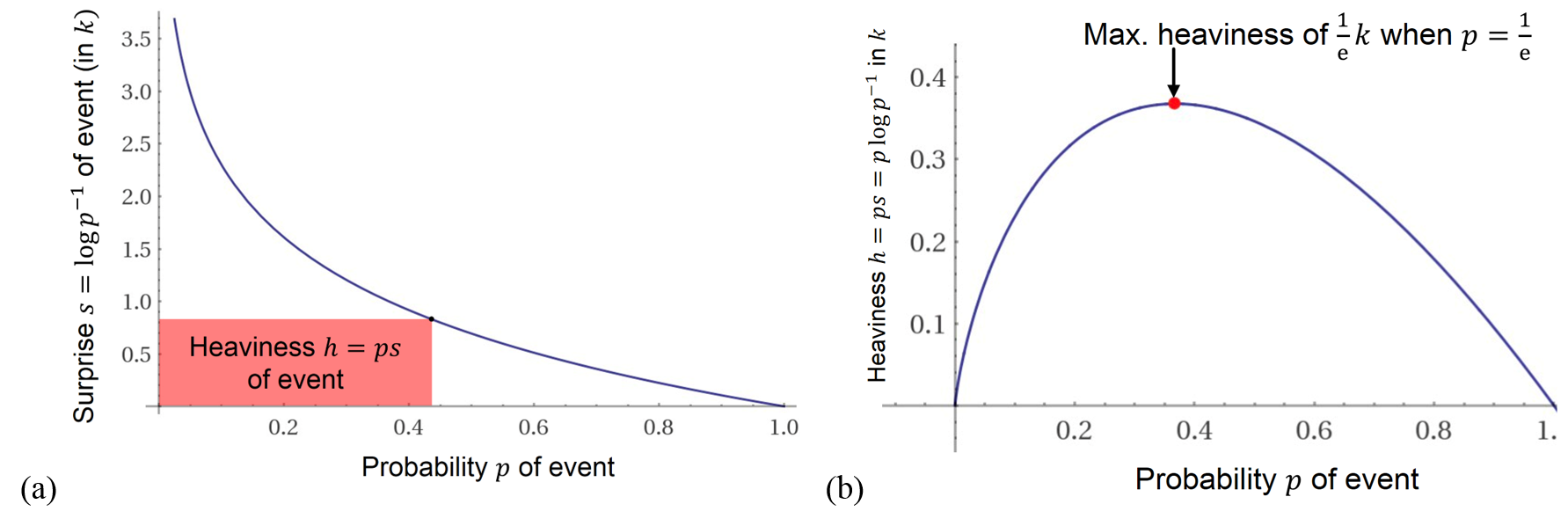}
        \caption{\textbf{\emph{Surprise and heaviness functions.}} 
            (a) Plot of surprise $s$ (in units of $k=\log \mathrm{e}$) as a 
            function of probability $p$.  Note that heaviness $h=ps$ is given by 
            the area of a rectangle drawn between the origin and a point on this 
            curve---if we imagine that the rectangle were a flat sheet of 
            physical material of uniform density and thickness, then its 
            physical heaviness would indeed be proportional to its area. (b) 
            Plot of heaviness (in $k$) as a function of probability.  Note that 
            the maximum heaviness of $k/\mathrm{e}$ is associated with events of 
            improbability e.}
		\label{fig:heavy}
	\end{figure}

Next, let's introduce a new technical concept that we call the {\emph{heaviness}} 
$h(v_i)$ of a value $v_i$, defined as its surprise $s_i=s(v_i)$, weighted by its 
probability $p_i=p(v_i)$ of occurring:  
    \begin{equation}
        h = h_i = h(v_i) = h(p_i) = p_i\cdot s_i = p_i\log m_i = -p_i\log p_i.
    \end{equation}
The heaviness function is plotted in Fig.~\ref{fig:heavy}(b).  Our use of the 
word ``heaviness'' for this concept is intended to evoke an intuitive 
psychological sense of the word, as in, how heavily does the possibility of this 
particular outcome weigh on one's mind?  The intuition here is that an extremely 
unlikely possibility doesn't (or shouldn't) weigh on our minds very heavily, and 
neither should an extremely likely one (since it is a foregone conclusion).  
This psychological interpretation of the concept will not be important to our 
later conclusions, though; it is merely provided to aid understanding.

It turns out that with the foregoing definition, the maximum heaviness inheres 
in an outcome that has an improbability of $m={\rm e}=2.71828\ldots$ (the base 
of the natural logarithms), or probability $p=1/{\rm e} = 0.3678879\ldots$; this 
carries a heaviness of $h=\mathrm{e}^{-1}\log {\rm e} = k\cdot 0.3678879\ldots,$ 
where $k$ is the logarithmic unit of knowledge defined by $k=\log \mathrm{e}$. 
(See Fig.~\ref{fig:heavy}(b).)  This logarithmic unit can also be identified 
with Boltzmann's constant $k_B$.  Whether the particular value e of the 
improbability at which ``peak psychological significance'' is supposedly 
attained in this conception could be substantiated by real psychological 
experiments is not important, however, to our present purposes; we are merely 
trying to instill some broad intuitive motivation here for these concepts.


\subsubsection{Expected value of a function.}
Next: Given any probability distribution $P=p(v_i)$ over a set $\mathbf{V} = 
\{v_i\}$ of values, and any function $f(v_i)$ of those values, we can define the 
{\emph{expected value of $f$ under $P$}}, written ${\rm E}_P[f]$, to be the sum 
of the $f(v_i)$ values weighted by their respective probabilities,
    \begin{equation}
        {\rm E}_P[f] = \sum_{i=1}^n p_i\cdot f(v_i).
    \end{equation}
This makes sense intuitively, since it is the (weighted) average value of the 
function $f$ that we would expect to obtain if values of $V$ were chosen at 
random in proportion to their probabilities.


\subsubsection{Entropy as expected surprise, or total heaviness.}
Now, for any probability distribution $P$ over any set of values $\mathbf{V} = 
\{v_i\}$, we can define the quantity called the {\emph{entropy}} of that 
distribution, as the {\emph{expected surprise}} $S=S(P)={\rm E}_P[s(p)]$ over 
all the different values $v_i\in\mathbf{V}$, or equivalently as the {\emph{total 
heaviness}} $H=H(P)$ of all the different values $v_i$:
    \begin{eqnarray}
        S(P) = \sum_{i=1}^n p(v_i)\cdot s(v_i) = \\
        H(P) = \sum_{i=1}^n h(v_i) = -\sum_{i=1}^n p(v_i)\cdot \log p(v_i).\label{eq:entropy}
    \end{eqnarray}

This statistical concept of entropy is, fundamentally, a property of an 
epistemological situation---namely, it quantifies how surprised we would expect 
to be by the actual value of the variable, if we were to learn it, or 
equivalently, how heavily our uncertainty concerning the actual value might 
weigh on our minds, if we dearly desired to know the value, but did not yet.  In 
simpler terms, we might say it corresponds to a {\emph{lack of knowledge}} or 
{\emph{amount of uncertainty}} or {\emph{amount of unknown information}}.  It is 
the extent to which our knowledge of the variable's value falls short of 
perfection.  We'll explain later why physical entropy is, in fact, the very same 
concept.

It is easy to show that the entropy $S(P)$ of a probability distribution $P$ 
over any given value set $\mathbf{V}$ has a maximum value of $S(P) = \hat{S}(V) 
= \hat{S}(\mathbf{V}) = \log n$ (where recall $n=|\mathbf{V}|$) when all of the 
probabilities $p_i$ are equal, corresponding to our original scenario, where 
only the number $n$ of alternative values is known.  In contrast, whenever the 
probability $p(v_i)$ of a single value $v_i$ approaches 1, the entropy of the 
whole probability distribution approaches its minimum of 0 (no lack of knowledge, 
{\ie} full knowledge of the variable's value).

We can also write $S(V)$ to denote the entropy $S(P)$ of a discrete variable $V$ 
under a probability distribution $P$ over the values of the variable that is 
implicit. 


\subsubsection{Conditional entropy.}
Another important entropy-related concept is {\emph{conditional entropy}}.  
Suppose that the values $v\in\mathbf{V}$ of a discrete variable $V$ can be 
identified with ordered pairs $(x,y)\in\mathbf{X}\times\mathbf{Y}$ of values of 
two respective discrete variables $X,Y$.  Then the {\emph{conditional entropy of 
$X$ given $Y$}}, written $H(X|Y)$, is given by
    \begin{equation}
        H(X|Y) = H(X,Y) - H(Y),
    \end{equation}
where $H(X,Y)=H(V)$ and $H(Y)$ is the entropy of the derived probability 
distribution $P(Y)=p(y_j)$ over ${\bf\rm Y}$ that is obtained by summing 
$P(V)=P(X,Y)$ (the joint probability distribution over all the ordered pairs 
$v=(x,y)$) over the possible values of $X$,
    \begin{equation}
        p(y_j) = \sum_{i=1}^{|\mathbf{X}|} p(x_i, y_j).
    \end{equation}

The conditional entropy of $X$ given $Y$ tells you the expected value of what 
the entropy of your resulting probability distribution over $X$ would become if 
you learned the value of $Y$.  That this is true is a rigorous theorem (which
we'll call {\emph{the conditional entropy theorem}}) that is provable from the 
definitions above.


\subsection{Some Basic Concepts of Information}
\label{sec:info}

In this subsection, we define and briefly discuss the quantitative concepts of 
{\emph{(known) information}}, {\emph{information capacity}}, and {\emph{mutual 
information}}.


\subsubsection{Known information: The complement of entropy.}
The amount of {\emph{information}} that is known about the value of a variable
is another statistical/e\-p\-i\-ste\-m\-o\-lo\-g\-i\-cal concept that is closely 
related to the concept of entropy that we just derived.  Entropy quantifies our 
{\emph{lack}} of knowledge about the value of a (discrete) variable, compared to 
the knowledge that we would expect to attain if the exact value of that variable 
were to be learned.  We just saw that the maximum possible entropy, in relation 
to a given discrete variable $V$ with a finite value set $\mathbf{V}$, is 
$\hat{S}(V) = \log |\mathbf{V}|,$ that is, the logarithm of the number of 
possible values of the variable, which is the same as the surprise that would 
result from learning the value, starting from no knowledge about the value.  
Thus, in any given epistemological situation (characterized by a probability 
distribution $P$) in which the entropy may be {\emph{less}} than that maximum, 
the natural definition of the {\emph{amount of knowledge}} that we have, or in 
other words the (amount of) {\emph{(known) information}} $K(P)=K(V)$ (also 
called {\emph{negentropy}} or {\emph{extropy}}) that we have about the value of 
the variable $V$, is simply given by the difference between the maximum entropy, 
and the actual entropy, given our probability distribution $P$:
    \begin{equation}
        K(P) = \hat{S}(\mathbf{V}) - S(P).
    \end{equation}

Note that we can also rearrange this expression as follows:
	\begin{eqnarray}
		K(P) &=& \hat{S}(\mathbf{V}) - S(P) 
						= \log n-\sum_{i=1}^n p_i \log \frac{1}{p_i} \\
		     &=& \sum_{i=1}^n p_i\log n + \sum_{i=1}^n p_i \log p_i 
		     			= \sum_{i=1}^n p_i(\log n + \log p_i) \\
		     &=& \sum_{i=1}^n p_i \log {np_i} 
		     			= \mathrm{E}_P[\log {np}] \\
		     &=& \mathrm{E}_P\left[\log \frac{p}{\overline{p}}\right] 
		     			= \mathrm{E}_P\left[\log \frac{\overline{m}}{m}\right],
		     			\label{eq:ip-vs-mp}
	\end{eqnarray}
where, in the last line (eq.~\ref{eq:ip-vs-mp}), we are referencing the baseline 
improbability $\overline{m}=n$ and baseline probability $\overline{p} = 
1/\overline{m}$ that we would have had in the default minimum-knowledge case.  
So, our knowledge or known information about a variable can be quantified as the 
expected logarithm of the multiplicative factor by which the probabilities of 
its outcomes are inflated (or improbabilities decreased), compared to the 
zero-information case.  

\subsubsection{Information capacity.} 
Clearly, the maximum knowable information $\hat{K}(V)$ about any variable $V$ is 
identical to its maximum entropy, $\hat{K}(V) = \hat{S}(V)$; we can also call 
this quantity the variable's total {\emph{information capacity}} $\mathbf{I}(V)$, 
and write
    \begin{equation}
        \mathbf{I}(V) = K(V) + S(V);\label{eq:info-capacity}
    \end{equation}
that is, in any given state of knowledge, the variable's total information 
capacity $\mathbf{I}$ (which is a constant) can be broken down into the known 
information $K$ about the variable, and the unknown information $S$ 
(entropy).\footnote{I gave a detailed example of this information capacity 
relation (eq.~\ref{eq:info-capacity}) in {\cite{Frank-02}}.}                              


\subsubsection{Mutual information shared between two variables.}
Next, given a situation with two discrete variables $X,Y$, with a state of 
knowledge about them characterized by a joint probability distribution $P(X,Y)$, 
the {\emph{mutual information between $X$ and $Y$}}, written $I(X;Y)$, is a 
symmetric function given by
    \begin{eqnarray}
        I(X;Y) = I(Y;X) &=& K(X,Y) - K(X) - K(Y),	\label{eq:mutinfo-K}	\\			
                        &=& H(X) + H(Y) - H(X,Y)		\label{eq:mutinfo-H}
    \end{eqnarray} 
in other words, it measures that part of our total knowledge $K(X,Y)$ about the 
joint distribution $P(X,Y)$ that is not reflected in the separate distributions 
$P(X)$ and $P(Y)$.  It is also the difference between the total entropies of 
(the probability distributions over) $X$ and $Y$ considered separately, and the 
entropy of the two variables considered jointly.  It is also a theorem that 
$I(X;Y)=H(X)-H(X|Y)$, the amount by which the entropy of $X$ would be reduced by 
learning $Y$ (and vice-versa).  Mutual information is always positive, and 
always less than or equal to the total known information $K(X,Y)$ in the joint 
distribution $P(X,Y)$ over the two variables $X,Y$ taken together.  It can be 
considered the amount of information that is {\emph{shared}} or redundant 
between variables $X$ and $Y$, in terms of our knowledge about them.  It can be 
considered to be a way of quantifying the degree of information-theoretic 
{\emph{correlation}} between two discrete variables (given a joint 
probability distribution over them).\footnote{Note that this 
        information-theoretic concept of correlation differs from, and is more 
        generally applicable than, a statistical {\emph{correlation coefficient}} 
        between scalar numeric variables.  General discrete variables do not 
        require any numerical interpretation.}


\subsection{Some Basic Concepts of Computation}
\label{sec:comp}

For our purposes in discussing {\lp}, it suffices to have an extremely simple 
model of what we mean by a (digital) computational process.  Our definition here 
will include stochastic (randomizing) computations, since these will allow us to
illustrate certain subtleties of the Principle.  The below definitions are 
essentially the same as the ones previously given in 
\cite{Frank-RC17,Frank-RC17-preprint,Frank-GRC18}.                                        


\subsubsection{Computational states and operations.}
Let there be a countable (usually finite) set $\mathbf{C} = \{c_i\}$ of distinct 
entities $c_i$ called {\emph{computational states}}.  Then a general definition 
of a (possibly stochastic) {\emph{(computational) operation}} $O$ is a function 
$O:\mathbf{C}\rightarrow \mathcal{P}(\mathbf{C})$, where $\mathcal{P}(\mathbf{C})$
denotes the set of probability distributions over $\mathbf{C}$.  That is, 
$O(c_i)$ for any given $c_i\in\mathbf{C}$ is some corresponding probability 
distribution $P_i : \mathbf{C} \rightarrow [0,1]$.  

The intent of this definition is that, when applied to an initial computational 
state $c_i$, the computational operation transforms it into a final 
computational state $c_j$, but in general, this process could be 
{\emph{stochastic}}, meaning that, for whatever reason, having complete 
knowledge of the initial state does not imply having complete knowledge of the 
final state.

Computational operations, under the above definition, can of course be composed 
with each other sequentially, to carry out a complex computational operation $O$ 
through a series of $\ell$ simpler steps, $O = O_\ell \circ O_{\ell-1} \circ 
\ldots \circ O_1$ (operating from right to left), but we will not delve into 
that aspect further here.


\subsubsection{Determinism and nondeterminism.}
For our purposes, we will say that a given computational operation $O$ is 
{\emph{deterministic}} if and only if all of its final-state distributions $P_i$ 
have zero entropy; otherwise we will say that it is {\emph{nondeterministic}} or 
{\emph{stochastic}}.  

The reader should note that this is a {\emph{different}} sense of the word 
``nondeterministic'' than the one most commonly used in theoretical computer 
science (\eg, in {\cite{Sipser-12}}).															


\subsubsection{Reversibility and irreversibility.}
We will say that an operation $O$ is (unconditionally logically) 
{\emph{reversible}} if and only if there is no state $c_k\in\mathbf{C}$ such 
that for two different $i,j$, $P_i(c_k)$ and $P_j(c_k)$ are both nonzero.  In 
other words, there are no two initial states $c_i$ and $c_j$ that could both 
possibly be transformed to the same final state $c_k$.  Operations that are not 
unconditionally logically reversible will be called {\emph{(logically) 
irreversible}}.

In \cite{Frank-RC17,Frank-RC17-preprint,Frank-GRC18}, we also defined a more              
general concept of {\emph{conditional}} logical reversibility, but for 
conciseness, we will not repeat that definition here.


\subsubsection{Computational scenarios.}
Finally, we can define a {\emph{computation}} or {\emph{computational scenario}} 
$\mathcal{C} = (O,P_\mathrm{I})$ as specifying {\emph{both}} a specific 
computational operation $O$ to be performed, and an initial probability 
distribution $P_\mathrm{I}$ over the computational state space $\mathbf{C}$.
We'll also refer to $P_\mathrm{I}$ as a {\emph{(statistical operating) context}}.  
Thus, a computational scenario, for our purposes, simply means that we have a 
(possibly uncertain) initial state $c_i$, and then we apply the computational 
operation $O$ to it.  It is easy to see that this then gives us the following 
probability distribution $P_\mathrm{F}$ over final states $c_j$:
    \begin{equation}
        P_\mathrm{F}(c_j) = 
            \sum_{i=1}^{|\mathbf{C}|} P_\mathrm{I}(c_i) \cdot P_i(c_j)
    \end{equation}
where $P_i = O(c_i)$ denotes the output distribution of $O$ for initial state 
$c_i$.

The above mathematical definitions regarding statistics, information and 
computation are now sufficient background to let us thoroughly explain the 
physical foundations of {\lp}.


\section{Information Theory and Physics}
\label{sec:infophys}

In this section, we discuss why the above information-theoretic concepts are 
appropriate and essential for understanding the role of information in modern 
physics, and specifically, the thermodynamics of computation.  As we will see, 
the absolute, rigorous correctness of {\lp} falls out as a direct consequence.


\subsection{The History of Entropy: from Clausius to Shannon}
\label{sec:ent-hist}

We begin by briefly reviewing the history of how the concept of {\emph{entropy}}
developed in physics; just knowing this history already illuminates why the
thermodynamic and information-theoretic concepts of entropy are not disparate,
but rather, are fundamentally interconnected.


\subsubsection{Clausius, 1850.}
When the concept of {\emph{entropy}} was first described, in a thermodynamic
context, by Rudolph Clausius in 1850 \cite{Clausius-1850}, its interpretation				
in terms of the above statistical definitions was not yet understood, and in 
fact, the information-theoretic quantity corresponding to entropy had not even 
been defined yet.  What Clausius noticed was that in any transfer of heat, a 
certain quantity $\Delt S = \Delt Q/T$, where $\Delt Q$ was the heat transferred 
in or out of a given system, and $T$ was the temperature of that system in 
absolute units, always was non-decreasing over time, when summed over all 
systems involved in the heat transfer.  The empirically-validated statement that 
total thermodynamic entropy is always non-decreasing is now known as {\emph{the 
Second Law of Thermodynamics}}.

The realization that physical entropy, which was originally described by 
Clausius as just a function of familiar thermal quantities such as heat and 
temperature, is actually also fundamentally a {\emph{statistical}} quantity, 
turns out to be a key part of the entire story of the subsequent progress of 
theoretical physics, as it advanced from classical mechanics to statistical 
mechanics and then to quantum mechanics.  This realization gradually took shape 
over several stages.


\subsubsection{Boltzmann, 1872.}
First, in the late 1800s, Ludwig Boltzmann began developing his theory of
statistical mechanics, in which he argued that the origin of familiar 
macroscopic thermal properties such as heat and temperature lay in the 
unobserved microscopic details of the mechanical behavior of the individual
particles (atoms and molecules) making up a given substance.  These particles
were too tiny and too numerous to observe or fully analyze their dynamics; they 
could only be treated statistically.  Moreover, the fundamentally discrete 
nature of physical states at the level of quantum mechanics was not yet known, 
so only continuous classical dynamics could be analyzed.  In his famous 
{\emph{H-theorem}} \cite{Boltzmann-1872}, Boltzmann defined a quantity he called 			
$H$ as (the negative of) what we would now call the entropy of a probability 
density function, which is the continuous analogue of a probability distribution 
over a discrete variable.  Boltzmann considered what would happen in a collision 
between two particles of an ideal gas if our initial knowledge about the 
positions of the particles consisted only of some probability density function 
representing a somewhat-uncertain initial position, and he found that the value 
of $H$ would in general become more negative as the particles interacted, 
corresponding to the knowledge of the state becoming more uncertain.  This 
analysis was an early illustration of the modern concept of {\emph{chaos}}, in 
which we find that the behavior of nonlinear dynamical systems generally tends 
to result in increased uncertainty about what their future state will be as we 
extrapolate farther into the future.  In any case, Boltzmann proposed that this 
increase of uncertainty about the detailed physical state of a system over time 
was the microscopic origin of the Second Law of Thermodynamics, and that thermal 
systems at equilibrium (or maximum entropy) were exactly those systems for which 
their probability density functions were already entirely spread out, 
corresponding to maximum statistical uncertainty ({\ie}, minimum knowledge) 
about their microscopic state.  However, prior to the development of quantum 
mechanics, microstates could not be described in terms of discrete variables, 
and so, without any way to count the number of microstates, the statistical 
basis for quantities such as the maximum entropy $\hat{S}$ of a system could not 
yet be made exact.


\subsubsection{Planck, 1901.}
In 1901, Max Planck \cite{Planck-1901} made Boltzmann's intuitions about the 		
statistical origin of entropy much more concrete when he analyzed the spectrum 
of blackbody radiation, and found that this spectrum could only be explained if 
electromagnetic energy could only exist in multiples of discrete {\emph{quanta}}
$E=h\nu$, where $h$ was a new constant (the quantum of action, what we now call
Planck's constant), and $\nu$ was the frequency of the radiation.  This 
discovery was the beginning of quantum mechanics.  Less widely known is that, as 
a side effect of his analysis, Planck also found that he could count the number 
of distinguishable microscopic states of an electromagnetic heat bath, and from 
this counting, derive (for the first time) an exact constant of proportionality 
$k$ between the classic thermodynamic entropy $S$ of a system at equilibrium, 
and the logarithm of the number of microscopic states, which, as suggested by 
Boltzmann's $H$-theorem, was the key quantity underlying thermodynamic entropy.  

Planck saw that, in the discrete case, a maximum statistical entropy $S=\hat{S}$ 
can be derived and expressed as 
\begin{equation}
S=k \log W,\label{eq:BP}
\end{equation}
where $W$ is the number of microstates, the logarithm here is base e by 
convention, and $k$ is the corresponding $\log\mathrm{e}$-sized unit of 
knowledge or entropy {\cite{Frank-arxiv-05}}, which (due to Planck's insight) 
can also be expressed in more conventional thermodynamic units of heat over 
temperature; this is the famous equation which ended up being carved on 
Boltzmann's tombstone to memorialize his role in the development of the 
statistical-mechanical concept of entropy.  But, it was actually {\emph{Planck}} 
who introduced the constant $k$ associated with the discreteness of states that 
is required to make Boltzmann's statistical entropy formula physically 
meaningful, and who first calculated the empirical value of $k$ in traditional 
thermodynamic units.  Planck's thermodynamic constant $k$ is what we now call 
{\emph{Boltzmann's constant}} $k_B$, to honor Boltzmann's role in laying the 
conceptual foundations of statistical mechanics.  

Thus, the origin of Boltzmann's constant and the origin of quantum mechanics are
{\emph{inextricably intertwined}} with each other; we could never have fully 
understood the statistical interpretation of entropy if we had not, at the same 
moment, understood that nature must be fundamentally discrete, and vice-versa.  
Quantum mechanics was {\emph{discovered}}, historically, as a direct logical 
consequence of using the statistical interpretation of thermodynamics to analyze 
the empirical blackbody spectrum.  Thus, you really can't believe in quantum 
mechanics without also believing in statistical mechanics, and vice-versa; the 
two theories {\emph{inherently}} go together.  And, these theories have been 
{\emph{enormously successful}; they comprise the conceptual foundations of almost 
all of the empirically-validated models of modern physics.  Arguably, rejecting the 
fundamental conceptual structure of these theories would be tantamount to rejecting 
almost all of the knowledge gained by 20th-century physics.

This is why we should be confident that any result (such as {\lp}) that follows 
logically from the most fundamental principles of statistical mechanics, such as 
Boltzmann and Planck's statistical interpretation of thermodynamic entropy, must
be correct.  At minimum, to coherently deny such a result would require finding 
an alternative conceptual framework (besides the foundation provided by standard 
statistical and quantum mechanics) that cogently explains all of the empirical 
results obtained by physics over the last 100+ years.  This seems highly unlikely,
which is why the recent experiments such as 
{\cite{Berut-etal-12,Orlov-etal-12,Jun-etal-14,Yan-etal-18}} are, from an					
historically-informed perspective, redundant with already-established science,
as far as proving {\lp} is concerned.  But, it's nevertheless a good thing that 
these experiments have been done, to help assuage the remaining skeptics.


\subsubsection{Von Neumann, 1927.}
Starting in the 1920s, von Neumann {\cite{VN27,VN32,VN55}} developed the                  
mathematical formulation of quantum mechanics that we still use today, and in 
the process, derived exactly how the Boltzmann-Planck concept of statistical 
entropy could be used to quantify uncertainty in quantum states; today, we call 
this quantum-mechanical formulation {\emph{von Neumann entropy}}, but it is 
essentially still just Boltzmann's original formulation of statistical entropy, 
as quantified by Planck.


\subsubsection{Shannon, 1948}
Finally, regarding the connection to the information-theoretic entropy that we 
already discussed in \S\ref{sec:stat}, which is usually described as having been 
formulated by Shannon \cite{Shannon-48,Shannon-49}:  Note, {\emph{Shannon cites 			
Boltzmann.}}  Effectively, {\emph{all}} that Shannon was really doing in 
defining his information-theoretic quantity $H$ was: (1) taking the statistical 
quantity $H$ that had {\emph{already been proposed by Boltzmann as the 
statistical meaning of thermodynamic entropy}},  (2) reversing its sign to match 
that of the thermodynamic entropy $S$, and (3) discretizing it to correspond to 
the discrete, quantum nature of reality that had already been discovered by 
Planck.  Further, after doing this, the formula for the entropy of a discrete 
probability distribution that Shannon ended up with was essentially identical to 
the one that had {\emph{already been developed more than twenty years prior} 
({\cite{VN27}}, reprised in {\cite{VN32,VN55}}) by von Neumann for use in 					
quantum mechanics.\footnote{A story is told, perhaps apocryphal, that at some
point von Neumann told Shannon that he should definitely call this concept 
entropy, because then no one would know what he was talking about!}  In other 
words, {\emph{Shannon was not introducing a new concept of entropy distinct from 
the existing one that was already being used in physics}}.  Rather, he was 
merely taking the {\emph{existing}} statistical concept of entropy that was by 
then already widely successful in physics, and simply applying it to the analysis 
of communication systems.

Moreover, Shannon himself explained, in the course of proving his channel 
capacity theorems, how the informational states of a digital communication 
system relate to distinct physical states {\cite{Shannon-49b}}; we will see that 			
this relation, which is well-validated by the empirical success of the modern 
communication systems which approach Shannon's bandwidth limits, is also central 
to the understanding of {\lp}.  In other words, {\emph{the underlying unity of 
information theory and statistical physics was an essential aspect of 
communication theory, from its very beginnings.}}  Communication theory could 
never possibly have been successful in engineering practice for optimizing the 
data rates of communication systems as a function of their physical parameters 
such as bandwidth and power levels, if this underlying unity had not been valid!  
Thus, information theory and statistical physics are most definitely {\emph{not}} 
unrelated domains of study that only coincidentally share some mathematical 
concepts, as certain critics of {\lp} have claimed.  That supposition is already 
belied by the practical success of communication theory.


\subsubsection{Conclusion of Historical Retrospective.}
Following Shannon, later authors such as Jaynes \cite{Jaynes-57} discussed 					
the connections between information theory and statistical mechanics in some 
depth, but such reviews should not even be necessary to explain the connection 
to those who already understand the above history of conceptual developments in 
statistical physics, and the essential role that the subject played in laying 
the intellectual foundations for Shannon's entire line of thought, and who know 
of the empirical success of information theory in engineering practice.

Thus, ``Boltzmann's constant'' $k_\mathrm{B}=k$ derives, at its root, from the 
statistical understanding of entropy and the quantum understanding of reality 
summed up in the Boltzmann-Planck formula (eq.~\ref{eq:BP}), and information 
theory itself (such as the basic definitions we reviewed in 
\S\S\ref{sec:stat}--\ref{sec:info}) is nothing but the language that was 
required to systematize and apply that foundation towards the engineering of 
physical artifacts that manipulate information; this includes computers as well 
as communication systems.

Further, all of the vast amount of 20th-century experimental physics that 
utilizes Boltzmann's constant also fundamentally rests (directly or indirectly) 
on the statistical-mechanical understanding of entropy.  Moreover, the entire 
structure of quantum theory rests, at its core, on the discreteness of states 
discovered by Planck, which itself was derived from statistical-mechanical 
assumptions.  Information theory is, fundamentally, {\emph{the}} basic language 
for quantifying knowledge and uncertainty in any statistically-described system, 
including physical systems.  And today's quantum physics is, at root, just the 
intellectual heir of Boltzmann's statistical physics, in its most 
highly-developed, modern form.  That's how deep the connection between 
information theory and physics goes.

The point of reviewing this history is simply to underscore this paper's main 
message, which is that to deny the validity of {\lp} would be to repudiate much 
of the progress in theoretical and applied physics that has been made in 
the more than 150 years that have elapsed since Boltzmann's earliest papers.


\subsection{Physical and Computational States}
\label{sec:states}

In this subsection, we review in some depth the relation between physical and 
computational states, as it has been understood since Shannon, and derive from 
it the equation relating computational and physical entropy, which we will call 
the {\emph{Fundamental Theorem of the Thermodynamics of Computation.}}


\subsubsection{Physical states.}
In the previous subsection, we recounted Planck's insight, which followed from 
his identification of the quantum of action $h$, that a given bounded 
thermodynamic system has only a countable, in fact finite, number of 
distinguishable microstates.  In modern quantum mechanics, the only refinement
to this insight of Planck's about the finiteness of the set of microstates is 
the realization that the physical state space can be broken down into 
distinguishable states in an uncountable infinity  of different ways---in 
technical terms, by selecting different orthonormal (mutually orthogonal and 
unit-normed) bases for the system's Hilbert 
space\footnote{A Hilbert space is a (typically) many-dimensional vector space 
        equipped with an inner product operator, defined over a field that is 
        usually the complex numbers $\mathbb{C}$.} 
of quantum state vectors.  Furthermore, the states can transform continuously 
into new states over time by rotating in this vector space, while maintaining 
the constraint that the number of distinguishable states at any given time 
remains constant and finite (for a finite system).  

Without delving into the full mathematical formulation of quantum mechanics, we 
can account for the key points for our purposes by simply stating that, for any 
quantum system with an $n$-dimensional Hilbert space, for any given time 
$t\in\mathbb{R}$, we will identify some set $\mathbf{\Phi}(t) = \{\phi_1(t),\, 
\phi_2(t),\, \ldots,\, \phi_n(t)\}$ of orthonormal vectors from that Hilbert 
space as ``the set of distinguishable microstates'' at time $t$.  An important 
point to know about quantum theory is that any uncertain quantum state (called a 
``mixed state'') can always be expressed as a simple probability distribution 
$p(\phi_i)$ over some appropriate basis set $\mathbf{\Phi}$.  The entropy of 
this probability distribution is called the von Neumann entropy of the mixed 
state (see \cite{VN27,VN32,VN55}), but it is the exact same 									
information-theoretic entropy quantity (for the given $p(\phi_i)$) that we have 
been referring to since \S\ref{sec:stat}.


\subsubsection{Computational states from physical states.}
Now, in relation to a typical real computer, the abstract computational states 
$c_i$ that we referred to in {\S\ref{sec:comp}} cannot necessarily be identified 
with uniquely-corresponding physical microstates $\phi_i$---since a general 
artifact intended as a ``computer'' will typically have many more possible 
microscopic variations in its physical structure (and the state of its 
surroundings) than computational states that it is designed to 
represent.  The only exception to this would be in the case of a 
conceptually-extreme quantum computer, in which every quantum number 
characterizing the configuration of the physical system making up its 
implementation---including, {\eg}, the spin orientation quantum number of every 
particle in the system---was considered as a part of its computational state.

In the more general case, there will be {\emph{a great many more physical states 
than computational states}}.  However, there clearly cannot be {\emph{fewer}} 
distinguishable physical states than computational states, since otherwise the 
computational states (when represented as physical states) would not be reliably 
distinguishable from each other, in violation of our assumption that they are 
distinct entities.  (For example, it would be physically impossible to reliably 
distinguish 3 different quantum state vectors selected from a 2-dimensional 
Hilbert space.)

However, there {\emph{is}} a definite relationship between computational states 
and physical states that {\emph{always}} holds, for {\emph{any}} real computing 
system:  Namely, {\emph{each well-defined computational state $c_i$ necessarily 
corresponds to a disjoint subset of some set $\mathbf{\Phi}$ of physical 
states}}.  (See Fig.~\ref{fig:states}.)  In other words, there is always some 
set $\mathbf{\Phi}$ of physical states, such that for each $c_i\in \mathbf{C}$, 
we can make the identification $c_i\subseteq\mathbf{\Phi}$, and for any two 
$i,j$, the subsets $c_i$ and $c_j$ do not overlap; $c_i\cap c_j = \varnothing$.  
We can also express this more concisely by saying that the set $\mathbf{C}$ of 
computational states is a (set-theoretic) {\emph{partition}} of some set 
$\mathbf{\Phi}$ of physical states, or (if not all physical states correspond to 
well-defined computational states) of one of its proper subsets.


    \begin{figure}[!tb]
        \centering
        \includegraphics[height=2in]{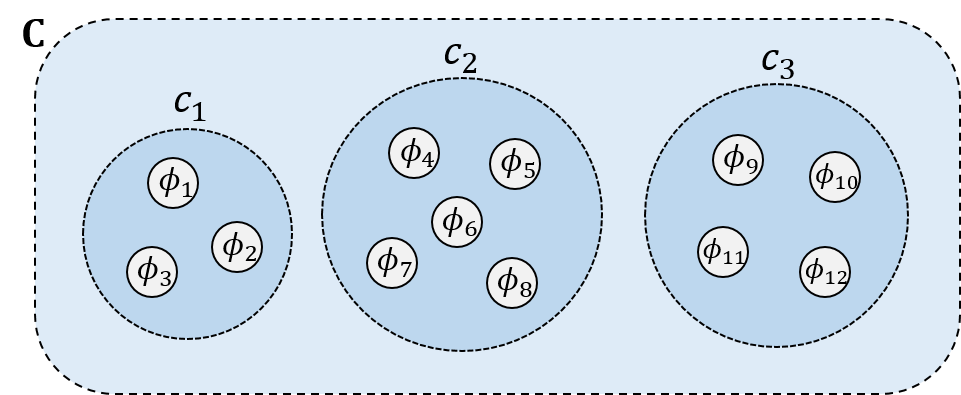}
        \caption{\textbf{\emph{Physical and computational states.}} 
            Example of a computational state space $\mathbf{C}=\{c_1,\, c_2,\, 
            c_3\}$ with 3 distinct computational states, where each state $c_i$ 
            is identified with a corresponding distinct subset $c_i = \Phi_i 
            \subseteq \mathbf{\Phi} = \{\phi_1,\, \phi_2,\, \ldots,\, 
            \phi_{12},\, \dots\}$ of a full set $\mathbf{\Phi}$ of all possible 
            physical microstates of the computer (or some larger physical 
            environment within which it is contained).  Typically in practice, 
            the number of distinguishable microstates per computational state 
            would be astronomically large.}
		\label{fig:states}
	\end{figure}


The reason why this must be the case is that, in order for a computational state 
$c_i$ to be well-defined, from a physical perspective, it must be possible, at 
least in principle, to reliably determine what the computational state 
{\emph{is}}, given some conceptually-possible measurement process ({\ie}, some 
quantum-mechanical {\emph{observable}} operator), which implies that there is 
some basis (implying a set of physical states $\mathbf{\Phi}$) for which, if we 
measure the state in that basis, we will obtain a definite answer (a specific 
physical state $\phi_j\in\mathbf{\Phi}$) that reliably reveals whether the 
computational state was $c_i$, or not.  Thus, the set of $\phi_j$ in this basis 
that would reliably imply that the computational state is $c_i$ may be 
identified with $c_i$.  This observation is very, very important:  It is why 
information entropy (in Shannon's conception) and physical entropy end up being 
fundamentally connected, as we will see.

Note that the above definition works even in the case of a quantum computer 
operating on any reliably-distinguishable set of input computational states, 
since even at any arbitrary point in the middle of a quantum computation, after 
some unitary time-evolution $U$ has been applied, there is always still some 
basis in which you could measure the computer's physical state such that, in 
principle, the original input state could be reliably determined (\eg, at 
minimum, in principle you could always apply $U^{-1}$ to get back to the initial 
state before doing the measurement).


\subsubsection{Computational and physical entropy.}
The above observation now lets us see why the information-theoretic entropy of a 
probability distribution over computational states is {\emph{necessarily}} 
fundamentally connected with physical entropy:  {\emph{Because the probability 
of a computational state is simply the sum of the probabilities of the 
corresponding physical microstates.}}  Let $P(c_j)$ denote the probability of 
the computational state $c_j$, and let $p_i = p(\phi_i)$ denote the probability 
of the physical state $\phi_i$.  Then we have:
    \begin{equation}
        P(c_j) = \sum_{\phi_i \in c_j} p_i.
    \end{equation}
Why {\emph{must}} this be the case?  {\emph{Because no other possibility is
epistemologically self-consistent.}}  Because, {\emph{given}} that the physical 
state is $\phi_i$, and that $\phi_i\in c_j$, it must be the case that the 
computational state is $c_j$, {\emph{by definition}}.  Thus, all of the 
probability mass associated with the physical states $\phi_i \in c_j$ 
contributes to the probability mass associated with $c_j$ (and nothing else 
does).

Now, the derived probability distribution $P(c_j)$ over the computational 
states $c_j$ implies a corresponding entropy $H(C)$ (the ``information entropy'') 
for the computational state $C$, considered as a discrete variable.  Similarly, 
the probability distribution $p(\phi_i)$ over the physical states $\phi_i$ 
implies a corresponding entropy $S(\Phi)$ (the ``physical entropy'') for the 
physical state $\Phi$, considered as a discrete variable.  {\emph{These two 
entropies necessarily have an exact and well-defined relationship to each 
other.}}  This is because the probability distribution $p(\phi_i)$ over the 
physical states {\emph{also acts as a joint distribution over the physical and 
computational states}}, because the computational state space is just a 
partition of the physical state space.\footnote{Even if not all physical states 
	correspond to well-defined computational states, we can always fix this by 
	simply adding an extra ``dummy'' computational state $c_0$ meaning, ``the 
	computational state is not well-defined.''}  
So, each physical state $\phi_i$ can thus also be identified with a pair 
$(\phi_i,c_j)$ of the values of these two discrete variables $\Phi,C$.  Thus, 
the conditional entropy theorem applies, and we can always write the following 
{\emph{Fundamental Theorem of the Thermodynamics of Computation}:
    \begin{equation}
        S(\Phi) = H(C) + S(\Phi|C).\label{eq:chain}
    \end{equation}
In other words, the (total) physical entropy $S(\Phi)$ is {\emph{exactly equal 
to}} the information entropy $H(C)$ of the computational state, plus the 
conditional entropy $S(\Phi|C)$ of the physical state, conditioned on the 
computational state---this just means, recall, the entropy that we would expect 
the physical state $\Phi$ to still have, if we were to learn the exact value of 
the computational state $C$.  This follows rigorously from the conditional 
entropy theorem (\ie, the derivation of the chain rule of conditional entropy).

As a convenient shorthand, we will call $S(\Phi|C)$ the {\emph{{\nc} entropy}} 
$S_\mathrm{nc}(\Phi)$ in contexts where the computational state variable $C$ is 
understood.  Thus, in such contexts, the Fundamental Theorem 
(eq.~\ref{eq:chain}) may also be written:
    \begin{equation}
        S(\Phi) = H(C) + S_\mathrm{nc}(\Phi).\label{eq:fundthm}
    \end{equation}

Another equivalent statement to eq.~\ref{eq:chain} is that $H(C) = I(\Phi;C)$, 
that is to say, the information entropy of the computational state is equal to 
the mutual information between the physical and computational state variables.  
This is obviously true, since the computational state can be thought of as being 
the state of an abstract physical system (``the computational system'') that is 
just {\emph{subsystem}} of the underlying physical system---so that, clearly, 
{\emph{all}} of our information about the computational system is redundant with 
our information about the physical system (since the computational system is 
just a {\emph{part}} of the physical system).

Simple as it is, we will call eq.~\ref{eq:chain} (or~\ref{eq:fundthm}) the 
{\textbf{\emph{Fundamental Theorem of the Thermodynamics of Computation}}, 
because essentially everything else that is important to understand about the 
subject rests upon it in some way.


\subsubsection{Visual proof of the Fundamental Theorem.}
\label{sec:vis}
Rather than reviewing the algebraic derivation that proves eq.~\ref{eq:chain} 
formally, we will describe  a simple visual representation of the theorem that 
makes plain why it is true.  This is where the {\emph{heaviness}} concept that 
we mentioned in {\S\ref{sec:stat}} becomes useful.  We saw in 
Fig.~\ref{fig:heavy}(a) that the heaviness or psychological weight of an outcome 
(value of a variable) can be visualized as a rectangle whose width is proportional 
to its probability, and whose height is proportional to its surprise or 
log-improbability.  Consider this rectangle, now, as one upwards-pointing branch 
of a tree, having one branch for each outcome.  The total heaviness of all the 
branches then corresponds to the entropy of the given probability distribution.

Thus, for example, in Fig.~\ref{fig:trees}(b), we see a tree representing a 
probability distribution over 5 physical states $\mathbf{\Phi} = \{\phi_1,\, 
\phi_2,\, \ldots,\, \phi_5\}$, where the probabilities are $p_1 = 1/12, p_2 = 
1/4, p_3 = 1/9, p_4 = 2/9, p_5 = 1/3$.  (The aspect ratio for the diagram is 
arbitrary, but the relative line heights and the relative line widths are 
otherwise to scale.)


    \begin{figure}[!tb]
        \centering
        \includegraphics[height=4in]{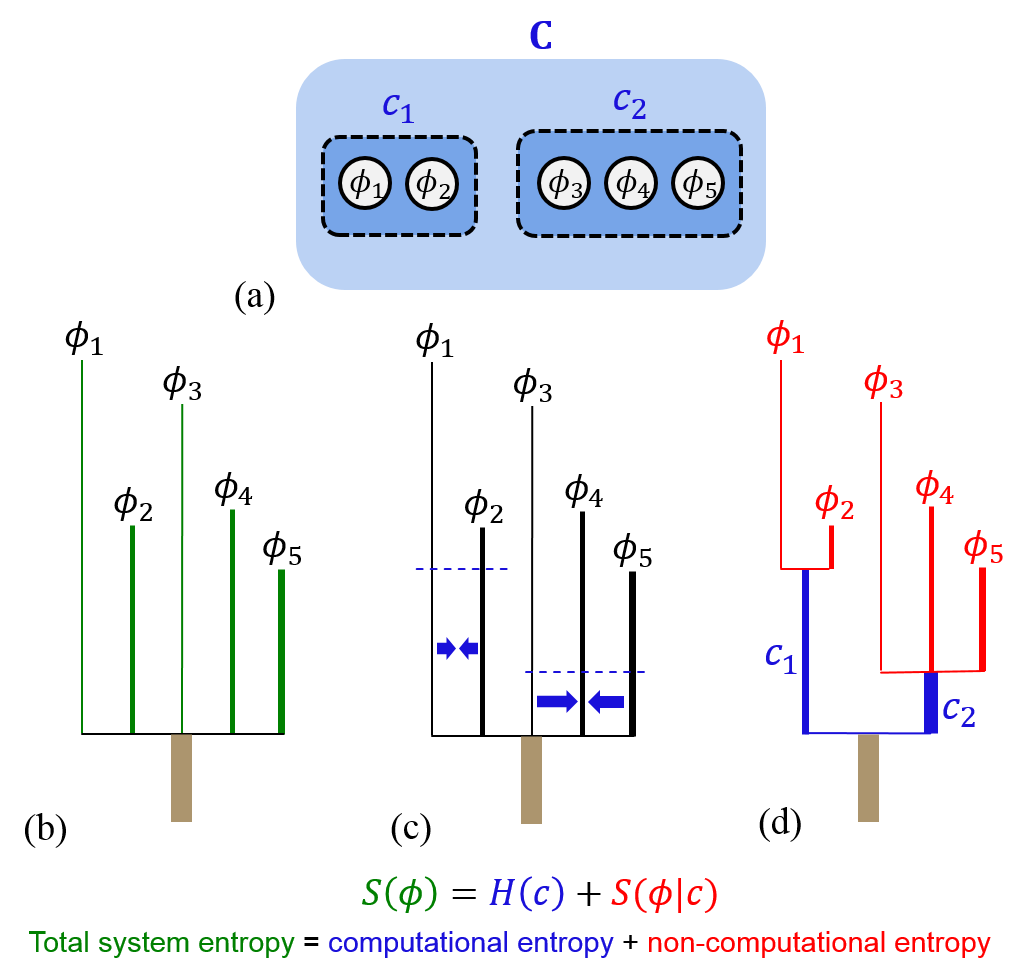}
        \caption{\textbf{\emph{Graphical illustration of the Fundamental Theorem 
            of the Thermodynamics of Computation.}} (a) Example of a 
            computational state space $\mathbf{C} = \{\,\{\phi_1,\phi_2\},\, 
            \{\phi_3,\phi_4,\phi_5\}\,\}$ constructed as a partition of a set 
            $\mathbf{\Phi}$ of 5 physical states $\{\phi_1,\, \ldots,\, \phi_5\}$. 
            (b) Tree representation of a probability distribution over $\phi_1,\, 
            \ldots,\, \phi_5$ given as 1/12, 1/4, 1/9, 2/9, 1/3. (c)-(d) Merging 
            of the lower parts of the branches to create ``trunk'' branches for 
            the computational states, and ``stem'' branches to represent the 
            conditional probability distribution over the physical states, given 
            the computational states.  As discussed in the text, it is easy to 
            see from the definition of conditional probability that the total 
            heaviness (area) of all branches remains the same before and after 
            the merge, and thus the Fundamental Theorem of the Thermodynamics of 
            Computation (eq.~\ref{eq:chain}) follows.}
        \label{fig:trees}
    \end{figure}


Now, if we wish to group individual outcomes into larger events corresponding 
to states of subsystems, like we do when we group physical states into 
computational states, we can represent this graphically by merging portions of 
branches into thicker branches.  So, for example, suppose that, as in 
Fig.~\ref{fig:trees}(a), the physical states $\{\phi_1,\, \phi_2\}$ are to be 
grouped into the computational state $c_1$, and the physical states $\{\phi_3,\, 
\phi_4,\, \phi_5\}$ are to be grouped into the computational state $c_2$.  Then 
we can use the derived probabilities $P(c_i)$ of the larger events $c_i$, 
together with the conditional probabilities $p(\phi_j|c_i) = p(\phi_j) / P(c_i)$ 
for the smaller events $\phi_j$, to create appropriate ``trunk'' (blue) and 
``stem'' (red) branches (see Fig.~\ref{fig:trees}(c,d)) for the micro-events 
$\phi_j$.  Note that the original probability is just the product of the new 
ones, $p(\phi_j) = P(c_i)\cdot p(\phi_j|c_i)$, and since the logarithm of a 
product is a sum, the length of the original branch is just the sum of the 
lengths of its corresponding trunk and the resulting stem.  In other words, the 
heights of all of the leaves of the tree are unchanged.  And since probabilities 
of mutually-exclusive sub-events add, the total width of each trunk is the same 
as the total width of the branches it is merged from.  So, it is easy to see 
visually that the total area or heaviness of this two-dimensional tree is the 
same after the merge.  Thus, the total entropy is the same.  Thus, the entropy 
of the computational state (blue) plus the entropy of the {\nc} state (red), or 
in other words the entropy of the physical state conditioned on the 
computational state, is the same as the total entropy of the physical state 
(green).  This is exactly what the Fundamental Theorem of the Thermodynamics of 
Computation is saying.

The appendix gives additional numerical and analytical details for this example.


\subsection{Physical Time-Evolution and Computational Operations}
\label{sec:time}

We now discuss how physical states dynamically evolve (transform to new states) 
over time, and relate this to our concept of computational operations from 
\S\ref{sec:comp}.  We begin by discussing how the law of non-decreasing entropy 
originally noticed by Clausius (the $2^\mathrm{nd}$ Law of Thermodynamics) 
follows as a direct logical consequence of the time-reversibility (injectivity) 
of microscopic dynamics.


\subsubsection{The reversibility of microphysics.} \label{sec:unitary}
For our purposes, the most important thing to know about the dynamical behavior 
of low-level physical states is that they evolve {\emph{reversibly}} (and 
deterministically), meaning, via bijective (one-to-one and onto) transformations 
of old state to new state.  

Formally, in quantum theory {\cite{VN32,VN55}}, over any time interval $\Delt t$,         
quantum states (mathematically represented as vectors in Hilbert space) are 
transformed to new state vectors by multiplying them by what in linear algebra 
are called {\emph{unitary}} matrices, {\ie} invertible linear operators that 
preserve vector norms (lengths).  Specifically, in any closed quantum system, the 
time-evolution operator is given by $U(\Delt t)=\mathrm{e}^{-\mathrm{i}\Delt 
tH/\hbar}$, where $\mathrm{i}=\sqrt{-1}$ is the imaginary unit, $\hbar=h/2\pi$ is 
the reduced Planck's constant, and $H$ is the {\emph{Hamiltonian}}, an Hermitian 
operator that is the total-energy observable of the system.  

For our purposes, the key point is that it is a mathematical property of unitary 
transformations that they preserve the {\emph{inner product}} between any two 
vectors (a complex analogue of a geometric dot product), which implies they 
preserve the {\emph{angle}} (in Hilbert space) between the vectors.  This is 
important because any two quantum state vectors $\ket{\psi_1},\, \ket{\psi_2}$ 
represent physically distinguishable states if and only if they are 
{\emph{orthogonal}} vectors, {\ie} at right angles to each other, meaning that 
their inner product $\braket{\psi_1 | \psi_2}=0$.  Thus, since unitary 
transformations preserve angles, {\emph{distinguishable quantum states always 
remain distinguishable over time}}.  So, if we identify our set of physical 
states $\{\phi_i\}$ with an orthonormal set $\{\ket{\psi_i}\}$ of quantum state 
vectors, it's guaranteed that these states transform one-to-one (injectively) 
onto a new set of mutually orthogonal states over any given time interval 
$\Delt t$.  

Setting aside the full linear algebraic machinery of quantum mechanics, 
we can summarize the important points about the situation for our purposes by 
saying that we have, for any given time $t\in\mathbb{R}$, a corresponding 
physical state space $\mathbf{\Phi}(t)$, such that, for any pair of times 
$s,t\in\mathbb{R}$, the dynamics among the states between these times is 
fully described by a total, bijective (one-to-one and onto) function $D(s,t) 
= D_{s}^{t}: \mathbf{\Phi}(s)\rightarrow\mathbf{\Phi}(t)$ mapping states at time 
$s$ to the states that they evolve to/from (depending on the sign of the time 
interval $\Delt t = t - s$) at $t$.  Further, for all $t\in\mathbb{R}$, $D_t^t$ 
is the identity function, and the dynamics is self-consistent, in the sense that 
for all $s,t,u\in\mathbb{R}$, $D_t^u \circ D_s^t = D_s^u$, {\ie}, the 
transformation that obtains from time $s$ to $t$, followed by the one from $t$ 
to $u$, is the same as the one from $s$ to $u$.


\subsubsection{The $2^\mathbf{nd}$ Law as a consequence of the reversibility of 
microphysics.}  As we mentioned briefly in 
\cite{Frank-RC17,Frank-RC17-preprint,Frank-GRC18}, it is easy to see that in any          
such {\emph{bijective dynamics}}, any initial probability distribution $P(t) = 
p(\phi_i(t))$ over the physical states at time $t$ will be transformed, over any 
time interval $\Delt t\in\mathbb{R}$, to what is essentially the same 
distribution over the corresponding new states, 
    \begin{equation}
        P(t+\Delt t) = p(\phi_i(t+\Delt t)) 
            = p\left(D_{t+\Delt t}^t(\phi_i(t+\Delt t))\right),
    \end{equation}
in other words, the probability of any state at time $t+\Delt t$ is identical 
to the probability of the state that it came from at time $t$.  Thus, the 
entropy $S(P)$ of the probability distribution is exactly preserved; $S(P(t_1)) 
= S(P(t_2))$ for all $t_1,t_2\in\mathbb{R}$.  So, when we know the precise 
microscopic dynamics $D$ and can exactly track its effects, entropy never 
increases or decreases (Fig.~\ref{fig:secondlaw}(a)).


    \begin{figure}[!tb]
        \centering\includegraphics[height=1.4in]{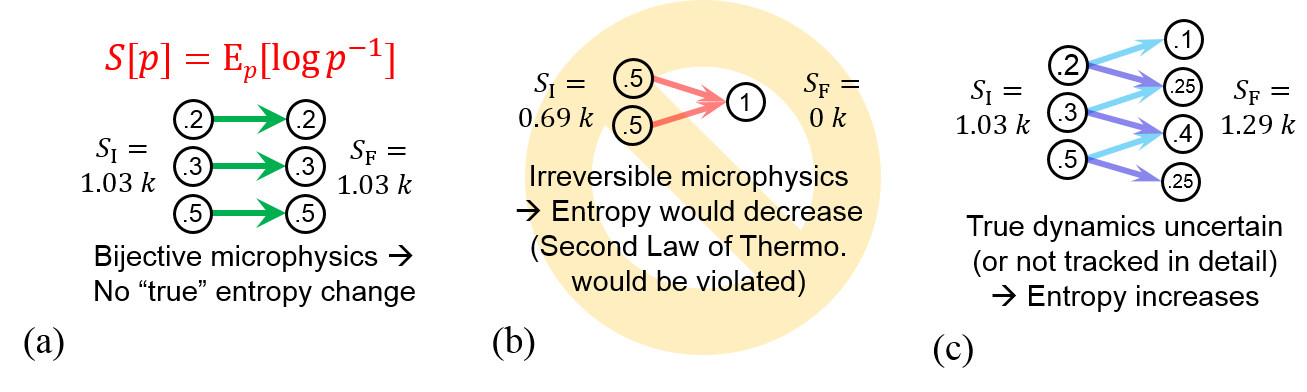}
        \caption{\textbf{\emph{The $2^\mathbf{nd}$ Law from bijective 
            microphysical dynamics.}} (a) Example of a bijective dynamics for a 
            simple system with 3 physical states.  The new states have identical 
            probabilities, and thus entropy is unchanged. (b) If the dynamics 
            was not injective, the $2^\mathrm{nd}$ Law of Thermodynamics would 
            be false.  If two states ever combined, then the illustrated initial 
            probability distribution would be changed to one of lower entropy.  
            This is true whenever the probabilities of two merged states are 
            both nonzero. (c) Entropy increases because we don't have exact 
            knowledge of the microscopic bijective dynamics, and/or don't have 
            the modeling capability to track its consequences in full detail, 
            so we replace the true dynamics with a stochastic one that 
            expresses our ignorance and/or incompetence.  In the illustration, 
            we treat the upwards-sloped and downwards-sloped injective 
            transformations as equally probable, resulting in a final 
            distribution that has greater entropy than the initial one.}
		\label{fig:secondlaw}
	\end{figure}


It is easy to see that the fact of the reversibility (bijectivity) of 
microphysics is actually a logical consequence of the Second Law of 
Thermodynamics (Fig.~\ref{fig:secondlaw}(b)), since if the dynamics $D$ was not 
always a one-to-one function, we would have two distinct physical states $\phi_1, 
\phi_2$ at some time $t$ that were both taken to the same state $\phi$ at some 
later time $t+\Delt t$ by the transformation $D_t^{t+\Delt t}$; their 
probabilities would be combined, and (it is easy to show), the heaviness 
(contribution to the total entropy) from the new state, $h(\phi)$, would 
necessarily be less than the sum of the heavinesses of the old states, 
$h(\phi_1)+h(\phi_2)$.  (This follows from the fact that the heaviness function 
is concave-down; see Fig.~\ref{fig:heavy}(b).)  Thus, total entropy would be 
decreased, and the Second Law would be false.  

How, then, can entropy increase?  Well, in practice, we do not know the entire 
dynamics $D$, or, even when we do, tracking its full consequences in microscopic 
detail would be beyond our capacity to accurately model.  If the dynamics $D$ is 
uncertain, or is simplified for modeling purposes by replacing it with a 
less-detailed model, then, even though we {\emph{know}} that the true underlying 
dynamics (whatever it is) must be one-to-one, the fact that in practice we have 
to replace the true dynamics with a statistical ensemble over possible future 
dynamical behaviors implies that, in this simplified model, the entropy will be 
seen as increasing.  This is illustrated in Fig.~\ref{fig:secondlaw}(c) for a 
simple case.  In this example, if the three states on the left (with 
probabilities 0.2, 0.3, 0.5) would transform bijectively to new states (on the 
right), but we had complete uncertainty about whether they would transform to 
the upper 3 states (upwards-sloping light blue arrows), or to the lower 3 states 
(downwards-sloping purple arrows), we would end up with a probability 
distribution over final states exhibiting greater entropy (in this case, by 
$0.26 k$) than the initial distribution.


\subsubsection{Computational operations and entropic dynamics.}
Let us now see what the bijective dynamics of microphysics implies about how 
entropy is transferred in computational operations.  First, we will expand our 
concept of a {\emph{computational state}} $c_i$ slightly, to account for the 
fact that the physical state space $\mathbf{\Phi}(t)$ will in general be 
changing over time, as individual states evolve according to the dynamics $D$.  
We will say that at any given time $t\in\mathbb{R}$, there is a computational 
state space $\mathbf{C}(t) = \{c_i(t)\}$ such that each computational state 
$c_i(t) \in \mathbf{C}(t)$ is a distinct subset of the physical state space 
$\mathbf{\Phi}(t)$ at that time, that is, $c_i(t) \subseteq \mathbf{\Phi}(t)$, 
and $c_i(t) \cap c_j(t) = \varnothing$ for all $i\neq j$.

Correspondingly, we must expand our notion of {\emph{applying a computational 
operation $O$}} in a computational scenario $\mathcal{C} = (O, P_\mathrm{I})$ to 
account for the fact that the computational states may be described differently, 
in terms of physical states, depending on exactly {\emph{when}} the operation 
starts and ends.  For this, we annotate the operation with its start and end 
times $s,t\in\mathbb{R}$, like $O_s^t$.  This notation then denotes that when 
the operation $O$ is applied from time $s$ to time $t$, the initial state 
$c_{\mathrm{I}i} = c_i(s)$ at time $s$ is mapped to final state $c_{\mathrm{F}j} 
= c_j(t)$ at time $t$ with probability $P_i(c_j) = O(c_i)(c_j)$, where here, 
$c_i,c_j$ label the time-independent computational states relative to which the 
original version of the operation $O$ was defined.

Now, let us examine more closely the consequences of applying a general 
computational operation $O_s^t$ from time $s$ to $t$, in the context of an 
underlying physical dynamics $D$ that is bijective.  

First, consider cases where $O$ is stochastic, so that there are computational 
state pairs $c_i,c_j$ such that $0 < P_i(c_j) < 1$; that is, a certain nonzero 
amount, but not all, of the probability mass from state $c_i$ at the initial 
time $s$ ends up in state $c_j$ at the final time $t$.  In order for this to be 
the case, when nothing is known about the initial physical state $\Phi(s)$ 
beyond what is implied by the initial computational state $C(s),$\footnote{\Ie, 
if $S(\, \Phi(s)\, |\, C(s)\, ) = \hat{S}(\,\Phi(s)\,|\,C(s)\,))$, or in other 
words, if $K(\Phi(s)) = K(C(s))$, so we have no more knowledge about the 
physical state than the computational state.} then $c_i(s)$ must correspond to a 
subset of $\mathbf{\Phi}(s)$ of initial physical states that itself has a proper 
subset $\Phi_i^j \subset c_i(s)$ consisting of states that will be mapped by the 
dynamics $D_s^t$ into the final state $c_j(t)$, and whose probability mass is a 
fraction $P_i(c_j)$ of the total.  Or, in equations,
    \begin{eqnarray}
        \Phi_i^j &=& 
            \{\;\;\phi_k(s) \in c_i(s)\;\;|\;\;D_s^t(\phi_k(s)) \in c_j(t)\;\;\}\\
		    \frac{|\Phi_i^j|}{|c_i(s)|} &=& P_i(c_j). \label{eq:stoch-frac}
    \end{eqnarray}
To explain eq.~\ref{eq:stoch-frac}, given a maximum-entropy conditional 
probability distribution $P(\Phi(s)\,|\,C(s))$, all of the microstates $\phi_k(s)$ 
in the given initial computational state $c_i(s)$ must be equally likely, so the 
ratio $|\Phi_i^j|/|c_i(s)|$ of the respective set cardinalities suffices to 
quantify $P(\phi \in \Phi_i^j\,|\,\phi \in c_i(s))$, the fraction of the total 
probability mass in $c_i(s)$ that is also in $\Phi_i^j$.  See 
Fig.~\ref{fig:stoch-frac} for an illustration.


    \begin{figure}[!tb]
        \centering
        \includegraphics[height=3.5in]{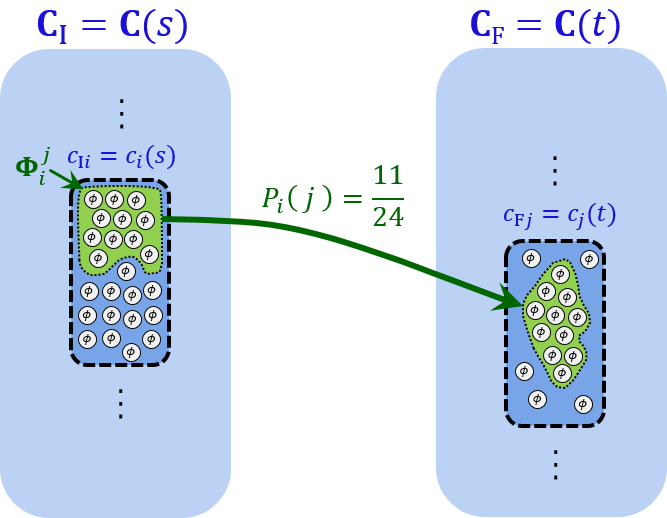}
        \caption{\textbf{\emph{Stochastic computation under bijective dynamics.}} 
            This diagram shows the relation of a stochastic computational 
            operation to bijective microphysics.  Illustrated are an initial 
            computational state space $\mathbf{C}_\mathrm{I} = \mathbf{C}(s)$ at 
            time $s\in\mathbb{R}$ and a final computational state space 
            $\mathbf{C}_\mathrm{F} = \mathbf{C}(t)$ at some later time $t > s.$ 
            Suppose that a stochastic computational operation $O_s^t$ is to be 
            performed such that the probability of going from some particular 
            initial computational state $c_{\mathrm{I}i} = c_i(s)$ at time $s$ to 
            the final state $c_{\mathrm{F}j} = c_j(t)$ at time $t$ should be 
            $P_i(j)=11/24=0.458\bar{3}$, and let the initial state of knowledge 
            be one in which the conditional probability distribution over the 
            initial physical state $\Phi(s)$ given the computational state 
            $C(s)$ is at maximum conditional entropy ({\ie}, the only 
            information known about the physical state $\Phi(s)$ is the mutual 
            information between the computational and physical state, which is 
            the information about the computational state, $K(\Phi(s)) = 
            I(\Phi(s); C(s)) = K(C(s))$).  Then it follows that all $\phi_k\in 
            c_j(t)$ are equally likely, and thus a fraction $11/24$ of these 
            physical states must be in the subset $\Phi_i^j \subset c_i(s)$ that 
            will be mapped into $c_j(t)$ by the micro-physical dynamics $D(s,t)$ 
            operating between times $s$ and $t$.  Note that here, 
            $c_{\mathrm{I}i}$ has only 24 microstates, and so exactly 11 of them 
            must go to $c_{\mathrm{F}j}$. More realistically, there would be an 
            astronomically-large number of microstates per computational state.}
        \label{fig:stoch-frac}
    \end{figure}


Finally, let's examine the entropic implications of performing an 
{\emph{irreversible}} computational operation $O_s^t$, which by definition means 
an operation in which some final computational state $c_{\mathrm{F}k}=c_k(t)$ at 
time $t$ has some nonzero probability of being reached from more than one 
initial computational state at time $s$, for example from both 
$c_{\mathrm{I}i}=c_i(s)$ and $c_{\mathrm{I}j}=c_j(s)$ for some $i\neq j$.  
Irreversible operations may generally reduce the entropy $H(C)$ of the 
computational state, as can be seen by setting the initial probabilities of both 
$c_{\mathrm{I}i}$ and $c_{\mathrm{I}j}$ to nonzero values (and all other 
initial-state probabilities to 0).  However, irreversible computational 
operations may still be implemented in bijective physics, but only by 
correspondingly {\emph{increasing}} the entropy $S_\mathrm{nc}(\Phi) = S(\Phi|C)$ 
of the {\nc} part of the state.  Why?  Because the Fundamental Theorem of the 
Thermodynamics of Computation (eq.~\ref{eq:fundthm}), together with the 
bijectivity of microphysics, ensures that the sum of computational and {\nc} 
entropies will be constant (or at least, non-decreasing, if the dynamics $D$ is 
uncertain).

For the case of a deterministic (non-stochastic) operation $O_s^t$, we can 
summarize the implications of the above observation very simply, by saying that 
between times $s$ and $t$, the required change (increase) $\Delt S_{\mathrm{nc}}$ 
in the {\nc} entropy $S_\mathrm{nc}(\Phi)$ of the physical state $\Phi$ is given 
by the negative of the (negative) change (decrease) $\Delt H(C)$ in the entropy 
of the computational state $C$ (the computational entropy); this is true in any 
statistical context, with any initial distribution $P_\mathrm{I}(C_\mathrm{I})$ 
over the initial computational state variable $C_\mathrm{I}=C(s)$:
	\begin{equation}
		\Delt S_\mathrm{nc}(\Phi) = -\Delt H(C) = H(C_\mathrm{I}) - H(C_\mathrm{F})
	\end{equation}

This observation is illustrated by the example in Fig.~\ref{fig:landauer}.


    \begin{figure}[!tb]
        \centering\includegraphics[height=4in]{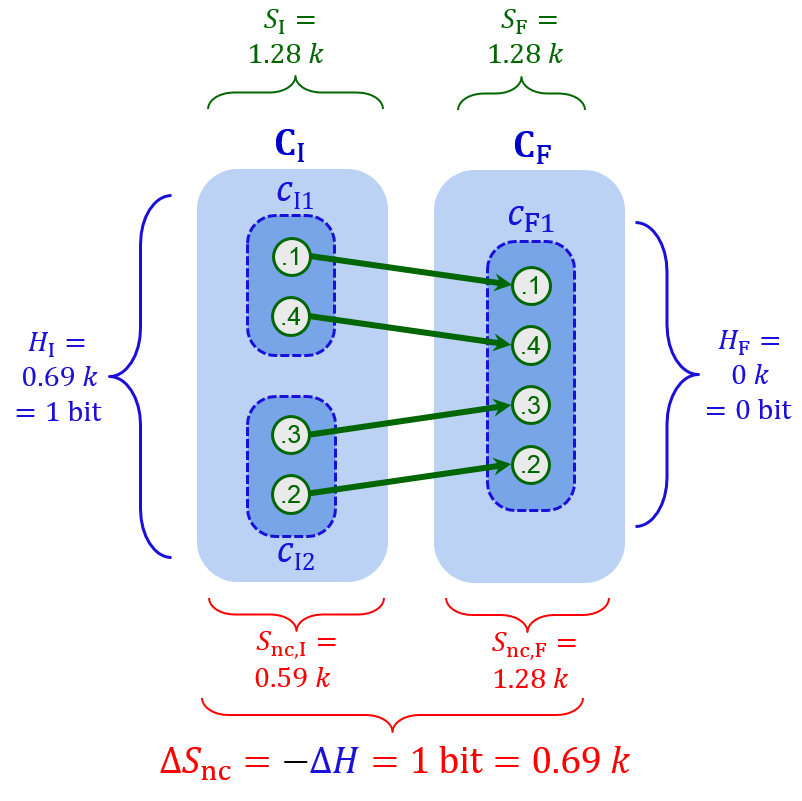}
        \caption{\textbf{\emph{Entropy ejection from the computational state.}} 
            Illustration of a deterministic, logically irreversible 
            computational operation in bijective microphysics.  Illustrated is 
            an initial computational state space $\mathbf{C}_\mathrm{I} = 
            \mathbf{C}(s)$ at time $s$ consisting of two possible initial 
            computational states $c_{\mathrm{I}1}=c_1(s)$ and $c_{\mathrm{I}2} = 
            c_2(s)$, and a final computational state space 
            $\mathbf{C}_\mathrm{F} = \mathbf{C}(t)$ at time $t>s$ consisting of 
            just one final computational state $c_{\mathrm{F}1}=c_1(t)$.  The 
            desired computational operation $O_s^t$ is one that maps both 
            $c_{\mathrm{I}1}$ and $c_{\mathrm{I}2}$ to $c_{\mathrm{F}1}$ with 
            certainty.  It follows from this that if the initial probability 
            distribution $P_{\mathrm{I}}(C_\mathrm{I})$ over the computational 
            states has some nonzero entropy $H(C_\mathrm{I})$, then the entropy 
            over the computational state will be reduced by an amount 
            $\Delt H(C) = -H(C_\mathrm{I})$, that is, to 0, and therefore (by 
            the Fundamental Theorem of the Thermodynamics of Computation, and 
            bijectivity), the entropy $S_{\mathrm{nc}}(\Phi)$ of the {\nc} state 
            will have to be increased correspondingly, {\ie}, $\Delt 
            S_{\mathrm{nc}}(\Phi) = -\Delt H(C) = H(C_\mathrm{I})$.  We can say 
            that all of the entropy in the computational subsystem has been 
            ejected into the {\nc} subsystem.  The figure shows state 
            probabilities for a case where the initial computational entropy is 
            $H_\mathrm{I}=H(C_\mathrm{I})=\log 2=1\,\mathrm{bit}\approx 0.69 k$, 
            and the initial {\nc} entropy was some arbitrary value (here about 
            $0.59 k$).}
        \label{fig:landauer}
    \end{figure}


\subsubsection{Intake of entropy by stochastic randomization.}
The above constitutes an important part of the argument for {\lp}.  However, 
this argument is not yet complete, for the following reason.  Processes such as 
the one illustrated in Fig.~\ref{fig:landauer} are actually 
{\emph{thermodynamically reversible}.  What do we get if we reverse in time a 
deterministic, logically irreversible process (by exchanging its initial and 
final times $s,t$)?  We exactly get a $\emph{stochastic, reversible}$ process,
which corresponds to performing a {\emph{measurement}} on the physical state.  
The time-reverse of Fig.~\ref{fig:landauer}, in particular, is a process that 
takes the final computational state $c_{\mathrm{F}1}$ stochastically back to 
either $c_{\mathrm{I1}}$ or $c_{\mathrm{I2}}$, with a probability distribution 
$P_i(j)$ that depends on the probability distribution over the physical states 
$\phi_k \in c_{\mathrm{F}1}$.  For a uniform (maximum-entropy) distribution over
physical states, the probabilities of returning to the initial states 
$c_{\mathrm{I1}}$ and $c_{\mathrm{I2}}$ would both be 0.5.  This is the same as 
the distribution we started with, so if we performed the process in 
Fig.~\ref{fig:landauer} forwards and then in reverse, the entropy of the 
computational state $H(C_\mathrm{I})$ would be unchanged.  However, if we 
allowed the physical states making up $c_{\mathrm{F}1}$ to be randomly 
reshuffled before the reversal, the final computational state might not be the 
same as the initial one.  Thus, such an operation (including the intermediate 
random permutation of the physical states) would be stochastic and logically 
irreversible, yet it could preserve the entropy $H(C_\mathrm{I})$ of the 
computational state overall.  (See Fig.~\ref{fig:revland}.)  It could also leave 
the {\nc} entropy $S_\mathrm{nc}(\Phi_\mathrm{I})$ of the physical state 
unchanged; for example, this would necessarily be the case whenever 
$S_\mathrm{nc}(\Phi)$ was already maximal initially, the initial and final 
computational entropies were maximal, and the detailed physical state was not 
further measured.


    \begin{figure}[!tb]
        \centering\includegraphics[height=1.5in]{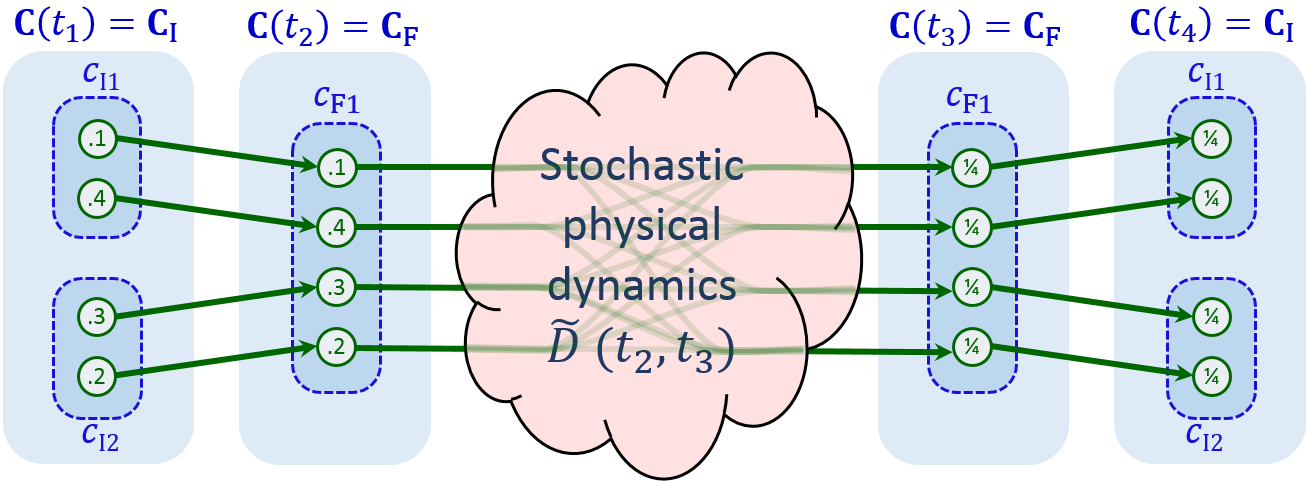}
        \caption{Illustration of conjoining a deterministic, logically 
        irreversible computational operation with its time-reverse, which is a 
        stochastic, logically reversible computational operation.  In between, 
        in this example we assume that a completely unknown physical dynamics 
        $\tilde{D}(t_2,t_3)$ occurs, which totally randomizes the physical 
        state, yielding a maximum-entropy distribution over the physical states 
        at time $t_3$.  In this example, the overall effect of the entire 
        process is that the entropy of the computational state remains unchanged 
        at $H(C)=1\,\mathrm{bit}$, and the entropy of the {\nc} state has been 
        increased from $\sim 0.85$ bits to 1 bit.  However, note that if the 
        initial {\nc} entropy had already been maximal (1 bit), then it could not 
        have increased further.  This illustrates that logically irreversible 
        operations on isolated, unknown computational bits do not necessarily 
        cause entropy increase, despite stochastic evolution of the environment; 
        the ejection of computational entropy to {\nc} form can sometimes be 
        undone by subsequent stochastic operations (measurements).  However, we 
        will later see that when logically irreversible operations are performed 
        in computational scenarios featuring multiple {\emph{correlated}} 
        computational state variables, the requirement for a permanent entropy 
        increase as per Landauer is recovered.}\label{fig:revland}
    \end{figure}


\subsubsection{Role of correlations.} 
Thus, entropy contained in {\emph{isolated, random}} computational bits, not 
having any correlations to any other available information, can be ejected to 
the environment in a {\emph{thermodynamically reversible}} way; another view of 
this process is illustrated in Fig.~\ref{fig:corr1}.  There, the 
merging/splitting of computational states is represented as an exchange of 
information between computational and {\nc} subsystems.  


    \begin{figure}[!tb]
        \centering\includegraphics[height=2.4in]{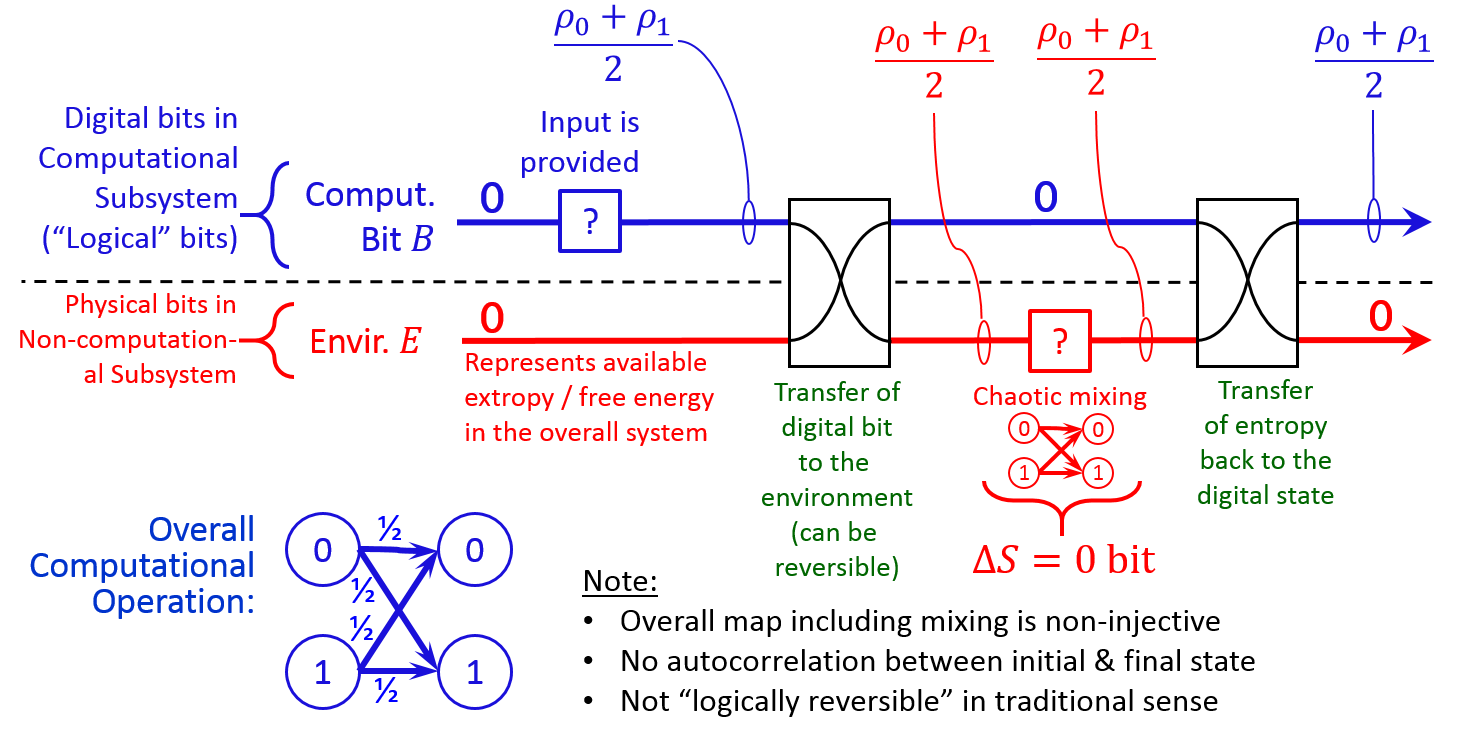}
        \caption{\textbf{\emph{Thermodynamically reversible erasure of an 
            unknown, uncorrelated bit.}}  Spacetime diagram showing an operation 
            sequence for the thermodynamically reversible erasure and 
            re-randomization of an isolated digital bit whose initial value is 
            unknown and uncorrelated with any other available information.  For 
            simplicity, we imagine that the computational and {\nc} subsystems 
            each have only 1 bit of information capacity (2 distinct states).  
            Initially, an input mechanism obtains some unknown bit-value from 
            the external environment, after which the computational bit $B$ has 
            a mixed state with 1 bit of entropy, representable by the density 
            matrix $(\rho_0 + \rho_1)/2$, where $\rho_i$ is a matrix 
            representing the state where bit $B$ has the unconditional value 
            $v_i=i$.  Suppose the environment bit $E$ is originally in a 
            ``cold,'' zero-entropy state.  We can reversibly swap bits $B$ and 
            $E$, moving the entropy from the computational subsystem to the 
            {\nc} one.  After this, the environment $E$ can undergo a stochastic 
            evolution that randomly scrambles its state---but this does not 
            increase its entropy, since it was already maximal.  Finally, we can 
            reversibly transfer the bit of entropy back to the digital state.  
            Overall, this process entails no net increase in entropy, yet is 
            logically irreversible, due to the stochastic evolution.}
        \label{fig:corr1}
    \end{figure}


However, in those examples, the fact that the digital bit that is being erased 
is initially uncorrelated with others is important.  Because the bit was 
uncorrelated with others, and its initial value was unknown, re-randomizing its 
value through the erasure/unerasure process does not actually decrease our known 
information, or increase entropy.  However, if the bit was initially 
{\emph{correlated}} with others, in the sense of sharing mutual information with 
them, then the situation is different.  This would be the case for any bits that 
are deterministically computed from others (see, {\eg}, Fig.~\ref{fig:corr2}).  
In this case, after the computed bit has been ejected to the environment, and is 
then randomized by the stochastic evolution of the environment, the prior 
correlation is lost, and total entropy is increased.  This consequence is truly 
unavoidable whenever we cannot track the exact microscopic dynamics of the 
environment, which is (by definition) always the case for a thermal environment, 
given that we do not have complete knowledge of (and capacity to keep track of) 
the microstate of the universe, nor do we know the complete laws of physics, the 
exact values of coupling constants, {\etc}


    \begin{figure}[!tb]
        \centering\includegraphics[height=2.25in]{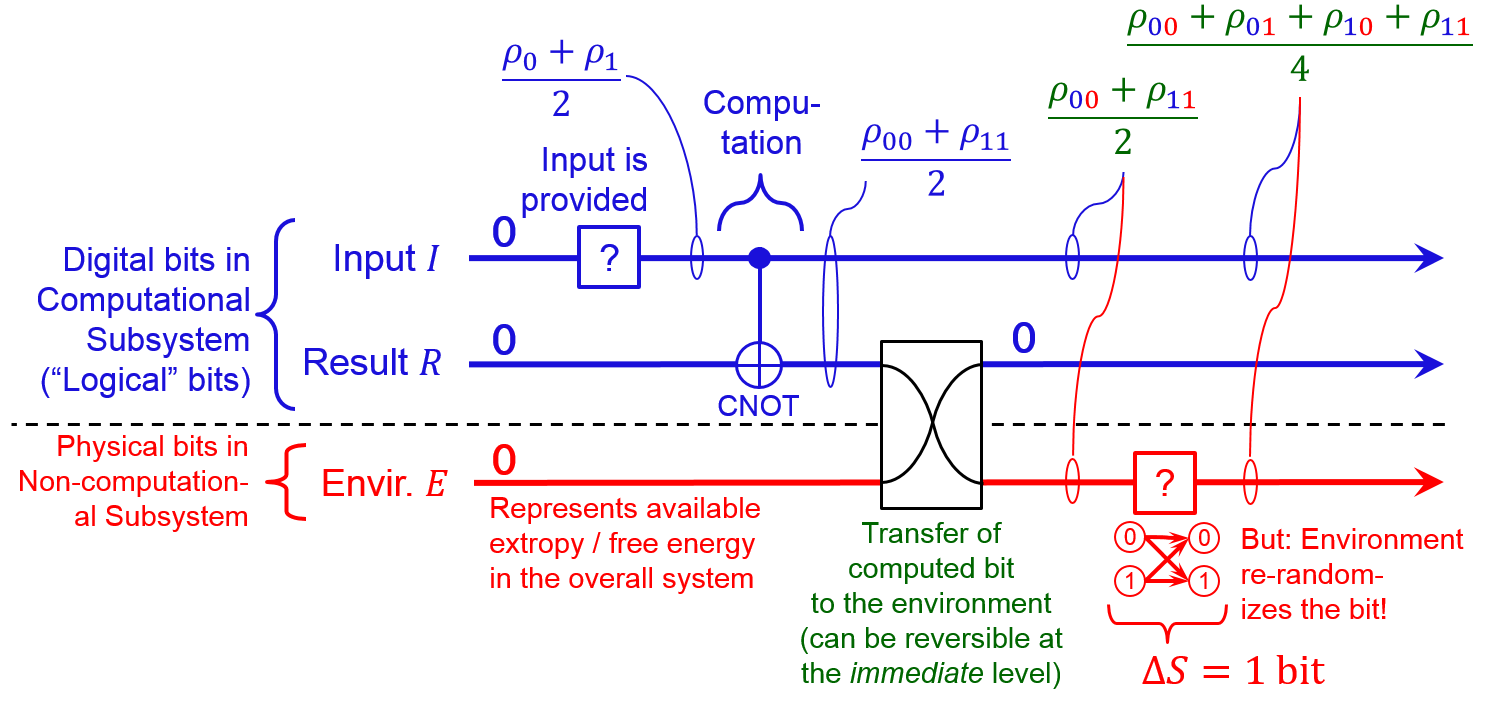}
        \caption{\textbf{\emph{Logically irreversible, oblivious erasure of a 
            correlated bit.}}  Spacetime diagram showing an operation sequence 
            for the thermodynamically {\emph{irreversible}}, {\emph{oblivious}} 
            erasure of a computed bit whose value is {\emph{correlated}} with 
            other available information.  Here, there are 2 bits $I,R$ in the 
            computational subsystem, and 1 bit $E$ in the {\nc} one.  As in 
            Fig.~\ref{fig:corr1}, an unknown input bit value is provided on the 
            input bit $I$, and $R$ is initially $v_0=0$, and then a 
            controlled-NOT operation is performed between $I$ and $R$. Now $I$ 
            and $R$ are {\emph{correlated}} (in the sense that they share 1 bit 
            of mutual information), and their joint mixed state can be 
            represented by the density matrix $(\rho_{00}+\rho_{11})/2$, where 
            $\rho_{ij}$ is a density matrix representing the state where $I=v_i$ 
            and $R=v_j$ with certainty.  Now, we can reversibly transfer one of 
            those bits $R$ as before, but now, when the environment 
            re-randomizes its bit $E$, this {\emph{loses the correlation}} 
            between $I$ and $E$, and our knowledge about the state is now 
            described by the density matrix $(\rho_{00} + \rho_{01} + \rho_{10} 
            + \rho_{11})/4$, which has 2 bits of entropy.  This represents a 
            permanent entropy increase of $\Delt S=1\,\mathrm{bit}$.  See also
            Fig.~\ref{fig:mutinfo}.}
        \label{fig:corr2}
    \end{figure}


A more detailed, fully general proof of {\lp} based on these observations about 
correlations goes as follows.  Let $X,Y$ be any two discrete random variables.  
For example, $Y$ could be a particular logical bit ({\ie}, a computational 
subsystem with a computational state space consisting of two distinct 
computational states) or set of bits of interest in a computer.  And $X$ could 
be the rest of the logical bits in the computer.

Now, given any joint probability distribution $P(X,Y)$ over these two variables, 
we know, as a matter of definition, that the mutual information between $X$ and
$Y$ is given by $I(X;Y) = H(X) + H(Y) - H(X,Y)$ (eq.~\ref{eq:mutinfo-H}), where
$H(X,Y)$ is just the usual von Neumann/Shannon entropy of the joint distribution
$P(X,Y)$, and where $H(X),H(Y)$ are just the usual (reduced) entropies of the
respective subsystems.

Note that whenever $I(X;Y) > 0$, a subsystem entropy value such as $H(Y)$ in
general {\emph{exaggerates}} the amount of {\emph{independent}} random 
information that is actually in $Y$, since part of the apparent uncertainty in
$Y$ is actually determined by (correlated with) $X$---namely, a part equal to
the mutual information $I(X;Y)$.

We can thus usefully define a quantity which we call the {\emph{independent
entropy of $Y$}} as
	\begin{eqnarray}
		S_{\mathrm{ind}}(Y) &=& H(Y) - I(X;Y)		\\
							 &=& H(Y|X),
	\end{eqnarray}
that is, it's just the conditional entropy of $Y$, conditioned on $X$.  (Recall,
this just means the expectation value of what the remaining entropy of $Y$
{\emph{would}} be, if we were to measure $X$ and learn its value.)

Suppose, now, that we erase $Y$ via a $local, oblivious$ mechanism---that is,
one that does not depend at all on the value of $X$ (or any other system that is
correlated with $Y$).  Typically, this could be done in a way that is completely
isolated from subsystem $X$, and does not interact at all with it.  We can do 
this erasure as slowly as we like.  Then, after waiting a bit (on a suitable 
thermalization timescale), we perform the reverse of this process, returning $Y$ 
to a state with the {\emph{same}} subsystem entropy $H(Y)$ as it had originally; 
note that this is the same case that we already exemplified in 
Fig.~\ref{fig:corr2} (with $X=I$ and $Y=R$), but in a more general context.

Note that, in this re-randomization process, it's now {\emph{impossible}} for
that process to restore any of the correlations with $X$, since we're not even
interacting with $X$ at all in the re-randomization process (since it is equally 
as oblivious to $X$ as the forward process was).

Another way to look at this is that, during the period when the correlated 
information that {\emph{was}} in $Y$ is instead out there in the 
{\emph{thermal}} (non-computational) environment, the mutual information that
the state originally had with subsystem $X$ is completely lost, it 
{\emph{vanishes}} (in the sense of, degrading to entropy) over the course of
that time (at or exceeding the thermalization timescale of the environment).

So in other words, at this point, despite the fact that $H(Y)$ is the same as
it was originally, now all correlations with $X$ have been lost, since the 
thermal environment (as per the very nature of what we mean by this term) won't 
have preserved those correlations in any accessible form---thus, now, $I(X;Y) = 
0$, and so $S_{\mathrm{ind}}(Y) = H(Y)$.  In other words, the independent 
entropy of $Y$ has now been increased, exactly by the amount $I(X;Y)$.

Note this implies a crucial observation: Whenever any subsystem $Y$ bearing any
nonzero amount of mutual information (shared with any other system $X$) is 
obliviously erased (without regards to $X$), this causes {\emph{an increase in 
the total entropy of the universe}} equal to (at minimum) the amount of mutual
information that $Y$ previously contained.

Now, suppose further that originally, $Y$ was, in fact, {\emph{deterministically
computed}} from $X$.  Note this is the case for {\emph{any}} bit in a computer, 
other than the input.  (Even for free memory, if we assume it has been 
initialized to a standard state, it can be considered just a constant function 
of the input.)

So then, since $Y$ is just a function of $X$, clearly, $H(Y|X) =0$ initially.  
And, $H(Y) = I(X;Y)$.  So, for example, if $Y$'s subsystem entropy is exactly 1 
bit, say ($\log 2 = k\ln 2$ entropy, meaning equal probability of 0 and 1), then 
so is its mutual information with $X$.

Thus, in such a case, erasing $Y$ (even quasistatically) and then reversing this
process results in a total entropy increase of $\Delta S = 1\,\mathrm{bit} = 
k\ln 2$, even though we have $H(Y) = 1$ at both the start and end of this 
process.  Because, the 1 bit's worth of correlation that it had with the rest of 
the system has been lost.  So, the new $P(X,Y)$ distribution has 1 more bit of 
entropy than it did previously.  And, there's been no decrease in environment
entropy to make up for this (because the erasure/restoration process was done
obliviously, it couldn't take advantage of the correlation to avoid 
increasing the entropy of the environment while part of the correlated state 
was being ejected).

Another way to describe the above process is to say that obliviously erasing 
computed bits turns their ``fake'' subsystem entropy ({\ie}, their mutual 
information that they have with other systems) into {\emph{real}} entropy, 
which is why total entropy increases.

The above argument is illustrated pictorically in Fig.~\ref{fig:mutinfo}, using
computational states illustrated as sets of physical states.


    \begin{figure}[!tb]
        \centering\includegraphics[height=1.25in]{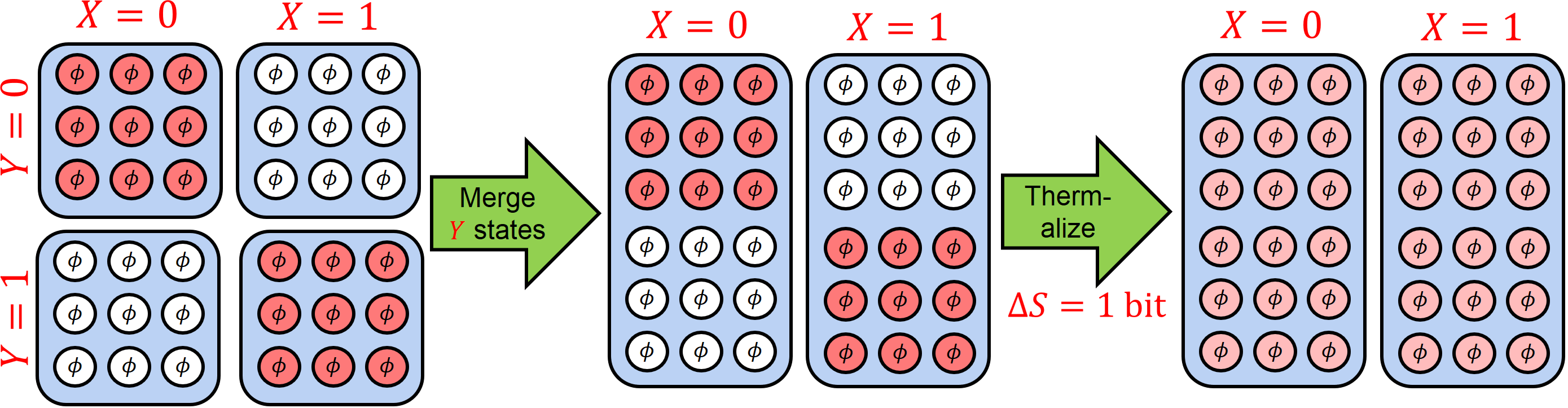}
        \caption{\textbf{\emph{Entropy increase from thermalization of mutual
        	 information.}}  (Left) Two perfectly-correlated computational bits $X$ 
        	 and $Y$; {\eg}, these could be the bits $I$ and $R$ from 
        	 Fig.~\ref{fig:corr2}.  (Middle) When $Y$ is obliviously erased, this 
        	 amounts to merging the two computational state spaces in each column; 
        	 we can say $Y=0$ in each merged space. Note that now, there briefly 
        	 exists a correlation between $X$ and the non-computational part of the 
        	 physical state.  (Right) Very quickly (over a thermalization 
        	 timescale), we lose track of the probabilities of the different 
        	 physical states making up each computational state, losing this 
        	 correlation.  This is where the absolute increase of total entropy 
        	 from {\lp} necessarily occurs.  We cannot then undo this entropy 
        	 increase by simply reversing the first step (un-merging the $Y$ 
        	 states), because the correlation information has already been 
        	 irrevocably lost by this point.}
        \label{fig:mutinfo}
    \end{figure}


Please note that the above argument is {\emph{absolutely}} mathematically
rigorous, and that it expresses the core essence of what is actually meant by 
{\lp}.  So, you really {\emph{can't}} deny Landauer if you understand basic 
math, the fact that information is conserved in physics (due to the 
$2^{\mathrm{nd}}$ Law of Thermodynamics and the unitarity of quantum mechanics, 
as we discussed in sec.~\ref{sec:unitary}), and you know what the concept of 
``thermalization'' means.

Incidentally, the above argument isn't novel, in the sense that, in its broad 
outlines, and/or at an intuitive level, it has already been well understood by 
myself and others in the thermodynamics of computing community for at least the 
last 20 years, if not longer.  In terms of explicit discussion of these ideas in 
the literature, our concept of ``independent entropy'' was previously called 
{\emph{non-information-bearing entropy}} by Anderson, as distinguished from 
{\emph{information-bearing entropy}} or mutual information; Anderson discusses 
these concepts, and the importance of correlations for understanding {\lp} and 
the thermodynamics of computation in a number of papers 
{\cite{Anderson-08,Anderson-14,Anderson-17,Anderson-18}}.									

\subsubsection{Reversible computing.}  Despite {\lp}, there is indeed a way 
in which correlations between bit values can be removed without increasing 
entropy, and that is precisely through {\emph{reversible computing}}; see 
Fig.~\ref{fig:corr3}.  In reversible computing, we take advantage of our 
knowledge of how a digital bit was computed to then reversibly 
{\emph{decompute}} it (\eg, via reversing the process by which it was computed 
originally), thereby unwinding its prior correlations, and restoring it to some 
known, standard, uncorrelated state which can then be utilized for subsequent 
computations.  In such a process, there is no need to transfer all or part of 
any correlated states to the {\nc} subsystem, which would cause those states to 
be randomized, and their correlations lost.  Thus, in contrast to the case 
illustrated in Fig.~\ref{fig:corr2}, there is no need for any entropy increase 
to result from a (generalized) logically reversible computational process, as 
we showed for the broadest class of deterministic classical computations in 
{\cite{Frank-RC17,Frank-RC17-preprint,Frank-GRC18}}.            						

Of course, various non-idealities present in our manufactured computational 
mechanisms in any given technology will generally result in some nonzero amount of 
entropy increase anyway, but that is a separate matter.  The key point is that 
{\emph{there is no known fundamental, technology-independent lower bound on the 
amount of entropy increase required to perform a reversible computation}}.  This 
sits in stark contrast to the case, in traditional irreversible computation, 
where we continually eject correlated bits to a randomizing environment; there, 
each bit's worth of correlated information that is lost in this way implies a 
$\log 2 = k_{\mathrm{B}} \ln 2$ amount of permanent entropy increase.  Thus,
reversible computing, if we continue to improve it over time, is indeed 
{\emph{the only physically possible way to perform general digital computation 
with potentially unlimited energy efficiency}}.


    \begin{figure}[!tb]
        \centering\includegraphics[height=2.75in]{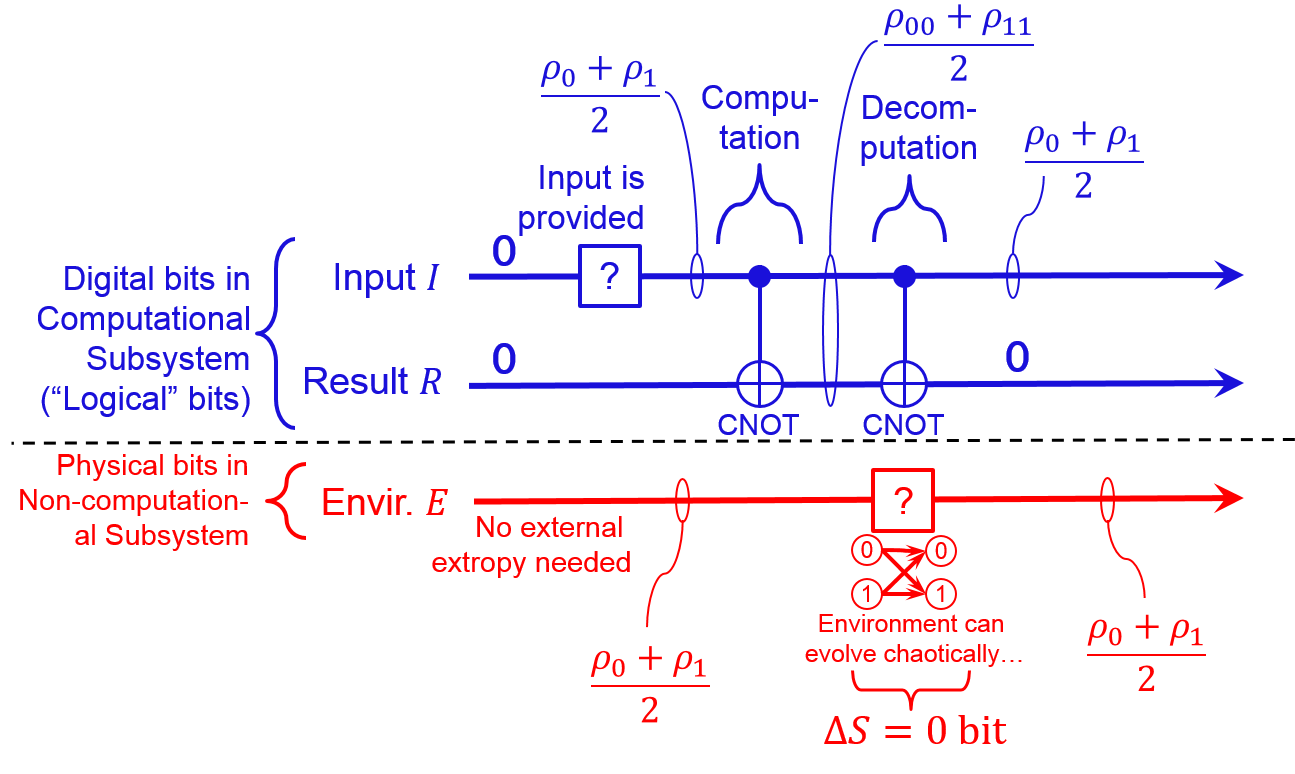}
        \caption{\textbf{\emph{Logically reversible, non-oblivious decomputation 
            of a correlated bit.}}  Spacetime diagram showing an operation 
            sequence for the thermodynamically and logically reversible, 
            non-oblivious {\emph{decomputation}} of a computed bit whose value 
            is correlated with other available information.  In this case, no 
            transfer of entropy is needed between computational and {\nc} states, 
            and the environment can start at maximum entropy.  As before, an 
            unknown input arrives on bit $I$, and then we XOR it into bit $R$.  
            But, rather than erasing $R$ by sending it to the environment, we 
            simply {\emph{decompute}} it in-place, by performing another CNOT 
            operation.  This removes the correlation between $I$ and $R$ 
            reversibly, and does not imply any increase in entropy.}
        \label{fig:corr3}
    \end{figure}


\subsection{Physical examples illustrating {\lp}.}
\label{sec:examples}

The above discussion of the rationale for {\lp} is at an abstract, albeit 
physically rigorous level.  In this section, we briefly describe a number of 
more concrete examples of physical systems that illustrate various aspects of 
the Principle that we have discussed.

\subsubsection{Bistable potential wells.}
One of the simplest systems that illustrates the points we've discussed is a 
bistable potential energy well with two degenerate ground states separated by 
a potential energy barrier (see Fig.~\ref{fig:potwell}).  This picture 
corresponds to a wide range of possible physical instantiations; {\eg} the 
wells could represent quantum dots, or states of certain superconducting 
circuits (such as parametric quantrons {\cite{Likharev-77}} or quantum flux               
parametrons \cite{Harada-etal-87,Hosoya-etal-91}), or ground states of many               
other systems.  These systems naturally support stable digital bits, encoded by 
the choice of which ground state the system is occupying at a given time.  The 
stored information has a lifetime corresponding to the timescale for tunneling 
of the system through the barrier, and/or thermal excitation over the barrier.  
(Which of these processes is dominant depends on the situation.)  At equilibrium, 
on sufficiently long timescales, the bit value will be unknown (entropy $\log 2$) 
and entropy of the system will not increase further, since it is already maximal.  
However, the bit's value at a given time (whatever it is) will be stable on 
shorter timescales; thus, this bit qualifies as a digital (computational) 
bit---{\eg}, it could be used for temporary storage in a computation.


    \begin{figure}[!tb]
        \centering\includegraphics[height=1.25in]{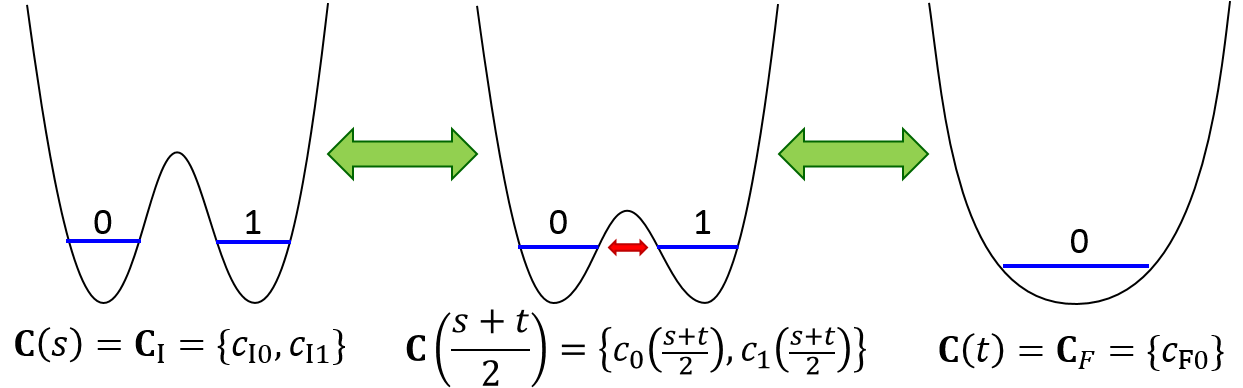}
        \caption{\textbf{\emph{Bistable potential well illustrating adiabatic 
            erasure/randomization.}}  This figure illustrates the 
            thermodynamically-reversible, logically-irreversible erasure (and 
            stochastic randomization) of an uncorrelated digital bit. (Left) 
            Consider a potential energy surface that includes two local minima 
            with a potential energy barrier between them---these could be, for 
            example, two adjacent quantum dots separated by a tunnel barrier. 
            Then, a subsystem that lives on that surface ({\eg} a surplus 
            electron in the quantum dot) will have two degenerate ground states, 
            one on each side of the barrier (we assume decoherence is 
            sufficiently strong in this system to prevent the stable ground 
            state from being a superposition of the two).  This can be a stable 
            digital bit, with a lifetime that corresponds to the tunneling 
            timescale.  (Center) However, if the height of the potential energy 
            barrier is lowered ({\eg}, by applying a suitable voltage to a gate 
            electrode above the tunnel barrier), the rate of tunneling between 
            the two states will increase, and the value of the bit will become 
            randomized on a shorter timescale.  (Right) Finally, if the barrier 
            is lowered completely, the two degenerate ground states will merge 
            into a single ground state, corresponding to an electron 
            wavefunction that straddles both dots.  Below the figures are 
            notations for the (time-dependent) computational state spaces, to 
            relate this picture to the theoretical discussion from earlier.  
            The corresponding physical state spaces will of course be much larger, 
            since they will include all of the microscopic thermal states of the 
            material and its environment, which (at nonzero temperature) will be 
            astronomically numerous.  Note, however, that if we adiabatically 
            transform the system from left to right and then back, the digital 
            state will be irreversibly randomized, but its entropy will not 
            increase if it was already maximal, and, in the adiabatic limit, 
            there will be no net increase in total system entropy.}
        \label{fig:potwell}
    \end{figure}

Now, consider what happens if we gradually lower the height of the potential 
energy barrier.  The rate of tunneling and/or thermal excitation over the 
barrier will increase, and the state will be randomized on ever-shorter 
timescales.  If we continue lowering the barrier to zero height, eventually we 
will be left with only a single stable ground state of the system.  This process 
corresponds to the process we've been describing, of pushing/ejecting a bit of 
computational information out to the {\nc} state of the environment.  If the 
digital state was initially known (correlated with other available information), 
then it is easy to see that this process results in a net entropy increase of $k 
\ln 2$ (as in Fig.~\ref{fig:corr2}).  The process of lowering the barrier can 
then be reversed, locking the system back into some stable digital state, but the 
bit value will have become randomized as a result, and our initial knowledge about 
its value, or any correlations, will be lost---entropy will have increased.  
However, if the initial digital state was {\emph{already}} unknown and 
uncorrelated, then its state information is {\emph{already}} all entropy, and so 
the process of lowering and re-raising the barrier does not need to increase 
entropy---in the adiabatic limit (if performed sufficiently slowly), the {\nc} 
entropy does not increase.  (But also, computational entropy has not decreased.)


\subsubsection{Adiabatic demagnetization.}
An example of a very well-studied physical phenomenon that illustrates the 
connection between information and physics is {\emph{adiabatic demagnetization}}, 
a.k.a. {\emph{paramagnetic cooling}} 
\cite{DeHaas-etal-34,Kunzler-etal-60,Pecharsky-etal-99}.  In this process,                
used in practice for certain cryogenic refrigeration applications, a magnetic 
biasing field is gradually removed from a sample of paramagnetic material, which 
allows the orientations of the magnetic dipoles in the material to randomize 
themselves.  In this process, entropy is transferred from the thermal, kinetic 
state of the material and its surroundings to the ``informational'' substrate of 
the dipole orientations.  Since neighboring dipoles tend to align, the dipoles 
will tend to cluster together into like-aligned domains of some size, which will 
be relatively stable; these are then very much in the nature of digital bits, 
and in fact, the information registered in the domains could be utilized in a 
computational process, as we do in magnetic media, such as disk drives or 
magnetic memory.  

Thus, adiabatic demagnetization is an example of a long-studied physical process 
by which {\nc} entropy (in the kinetic, thermal state of a system) can be 
transferred to the form of what is effectively digital, computational entropy, 
and the thermodynamic impact of this transfer of entropy to this more obviously 
``informational'' form is directly measurable.  Similarly, the process can be 
reversed, by gradually applying a field to re-align the dipoles, ``erasing'' 
their digital content and thereby heating up their surroundings.  The results of 
all the many decades of laboratory experiments performed on these processes are 
exactly consistent with standard statistical mechanics, and the entire view of 
the thermodynamics of computation that we have been discussing.


\section{Empirical studies validating {\lp}}
\label{sec:experiments}

{\lp}, as explained above, directly and rigorously follows from the enormous 
and sophisticated success of the theoretical understanding and empirical 
validation of the concepts of statistical physics that has obtained over the 
century and a half that have passed since Boltzmann's pioneering insights.  
But if any additional assurances are needed, there have been several experiments 
in recent years that have demonstrated the correctness of {\lp} more directly.  
Here we review a few of those experiments, very briefly.

In 2012, Berut {\etal} {\cite{Berut-etal-12} tested {\lp} in the context of a             
colloidal particle trapped in a modulated double-well potential, an experimental 
setup designed to mimic the conceptual picture that we reviewed in 
Fig.~\ref{fig:potwell}.  Their experimental results showed that the heat 
dissipated in the erasure operation indeed approached the Landauer value of 
$k_\mathrm{B}T\ln 2$ in the adiabatic limit.  Also in 2012, Orlov {\etal} 
{\cite{Orlov-etal-12}} tested {\lp} in the context of an adiabatic charge                 
transfer across a resistor, and verified that, in cases where the charge 
transfer is carried out in a way that does not erase known computational 
information, the energy dissipated can be much less than $k_\mathrm{B}T\ln 2$, 
which validates the theoretical rationale for doing reversible computing.  In 
2014, Jun {\etal} {\cite{Jun-etal-14}} carried an even more high-precision                
version of the Berut experiment, verifying again the Landauer limit, and that 
similar, logically-reversible operations can, in contrast, be done in a way that 
approaches thermodynamic reversibility.  Finally, in 2018, Yan {\etal} 
{\cite{Yan-etal-18}} carried out a quantum-mechanical experiment demonstrating            
that {\lp} holds at the single-atom level.

In contrast, the only experiments that have claimed to demonstrate violations of 
Landauer's limit have been ones in which the experimenters misunderstood some 
basic aspect of the Principle, such as the need to properly generalize the 
definition of logical reversibility, which was the subject of 
{\cite{Frank-RC17,Frank-RC17-preprint,Frank-GRC18}}, or the role of correlations          
that we explained in \S\ref{sec:time} above.


\section{Conclusion}
\label{sec:conclusion}

In this paper, we reviewed a number of aspects of {\lp}, including its 
historical origin in the very foundations of statistical physics, which laid the 
essential groundwork for modern statistical thermodynamics and quantum mechanics.  
We saw that information theory is perfectly suited to examining the role of 
information in physics, and in fact its development historically {\emph{grew out 
of statistical physics}}.  Then we detailed exactly how the high-level view of 
information and computational operations in any real computer connects 
fundamentally (and unavoidably) with the physical concepts of distinguishable 
states and bijective dynamics that are essential features of all modern ({\ie}, 
quantum) models of fundamental physics.  We explained exactly why an 
irreversible, permanent increase in entropy of $\log 2 = k \ln 2$ upon the 
logically-irreversible, oblivious erasure of a correlated bit is an 
unavoidable, and {\emph{totally mathematically rigorous}} consequence of these 
fundamental physical theories, and why, in contrast, a reversible computational 
process can {\emph{completely avoid}} the resulting Landauer limit on the 
energy efficiency of computation, something that traditional computational 
mechanisms, which discard correlated bits every time a logic gate destructively 
overwrites its previous output, can never do.  Therefore, as reversible 
computing technologies continue to be improved over time, they can potentially, 
in the long term, become {\emph{unboundedly more energy-efficient}} than 
{\emph{all physically possible irreversible computers}}.  Meanwhile, the 
correctness of {\lp}, and the fact that only reversible computational processes 
can circumvent it, have already been directly empirically validated in various 
experiments.

One caveat to the above statements that could use some further elaboration comes 
from our observation in \S\ref{sec:time} (also mentioned in {\cite{Frank-05}}) 	          
that isolated digital bits that are {\emph{already}} entropy ({\ie}, uncorrelated 
with any other bits, and unobserved) can be re-randomized, either in-place, such 
as when the potential barrier is partially lowered in Fig.~\ref{fig:potwell}, or 
by ejecting them to a thermal environment, allowing the environment to randomize 
them, and subsequently taking them back in, like in adiabatic 
magnetization/demagnetization, without a necessary increase in total entropy.  
This raises some potentially interesting algorithmic possibilities for performing 
randomized computations more energy-efficiently (and securely).  For example, 
cryp\-to\-graph\-i\-cal\-ly-secure random bits can be taken in by transferring 
their entropy adiabatically from a thermal environment, after which a 
probabilistic algorithm can be executed (reversibly) using those bits, and then 
(after results are obtained) the utilized entropy can be re-isolated by 
reversing the computation, after which the random bits used can be pushed back 
out to the thermal environment, thereby losing them permanently (giving a 
forward secrecy property) as the environment re-randomizes them, with 
asymptotically zero net new entropy having been generated in this entire process.  

Somewhat more generally, we can also develop a more comprehensive theoretical 
treatment of the thermodynamics of stochastic computational operations.  We 
could extend the theoretical tools presented in 
{\cite{Frank-RC17,Frank-RC17-preprint,Frank-GRC18}} and the present paper, to             
derive the thermodynamic implications of performing arbitrary, general 
computational operations in statistical contexts featuring any arbitrary initial 
probability distributions.  This would include stochastic and irreversible 
operations performed in contexts that exhibit varying degrees of correlation 
between the part of the state that is being manipulated by the operation, and 
other parts of the computer.  It is conceivable that in the course of 
undertaking such an investigation, we might uncover a few less-obvious 
algorithmic opportunities.  Developing this more general theory is beyond the 
scope of the present paper, but would be an appropriate target for future work.

It is hoped that the present paper will help to clarify the fundamental physical 
justification of {\lp}.  This is especially important since the possibility of 
approaching reversible computation presents us with the opportunity to 
eventually make unboundedly-greater gains in the amount of economic value that 
our civilization may extract in the future from any given energy resources via 
computation, compared to the best that we could ever accomplish without it.  I 
encourage readers armed with this understanding to invest their own personal 
energies into helping to develop the reversible computing paradigm towards 
practical application in the engineering of more efficient computers.


\newpage


\section*{Appendix: Numerical example of the main theorem}

In \S\ref{sec:vis}, we described a visualization method to facilitate 
understanding of the Fundamental Theorem of the Thermodynamics of Computation
(eq.~\ref{eq:chain}).  The following illustrations 
(Figs.~\ref{fig:tree1}--\ref{fig:tree3}) give more detailed numerical values 
(and some related analytical formulas) for the example summarized in 
Fig.~\ref{fig:trees}.  Note that here, the trees are displayed
sideways to save space.  Surprisingnesses (horizontal scale) and 
heavinesses/entropies (proportional to individual/total branch areas) are all 
dimensioned in units of $k=\log \mathrm{e}$, equivalent to Boltzmann's 
constant.  The branch thicknesses (vertical scale) are proportional to their 
(dimensionless) probabilities.


	\begin{figure}
		\centering\includegraphics[height=2.5in]{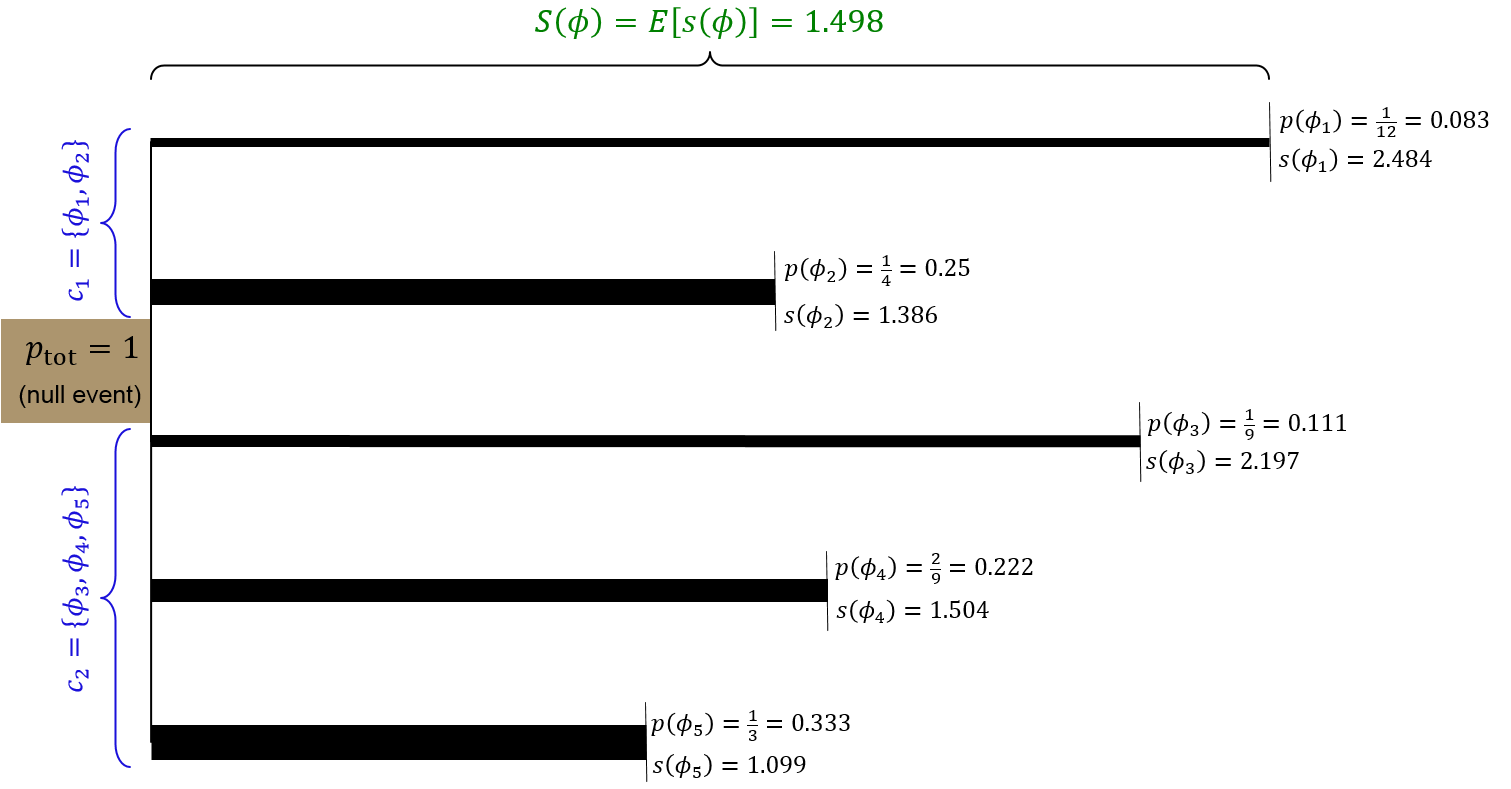}
		\caption{Detailed numerical data for the example of 
		Fig.~\ref{fig:trees}(a), before the grouping operation.}
		\label{fig:tree1}
	\end{figure}
	

	\begin{figure}
		\centering\includegraphics[height=2.75in]{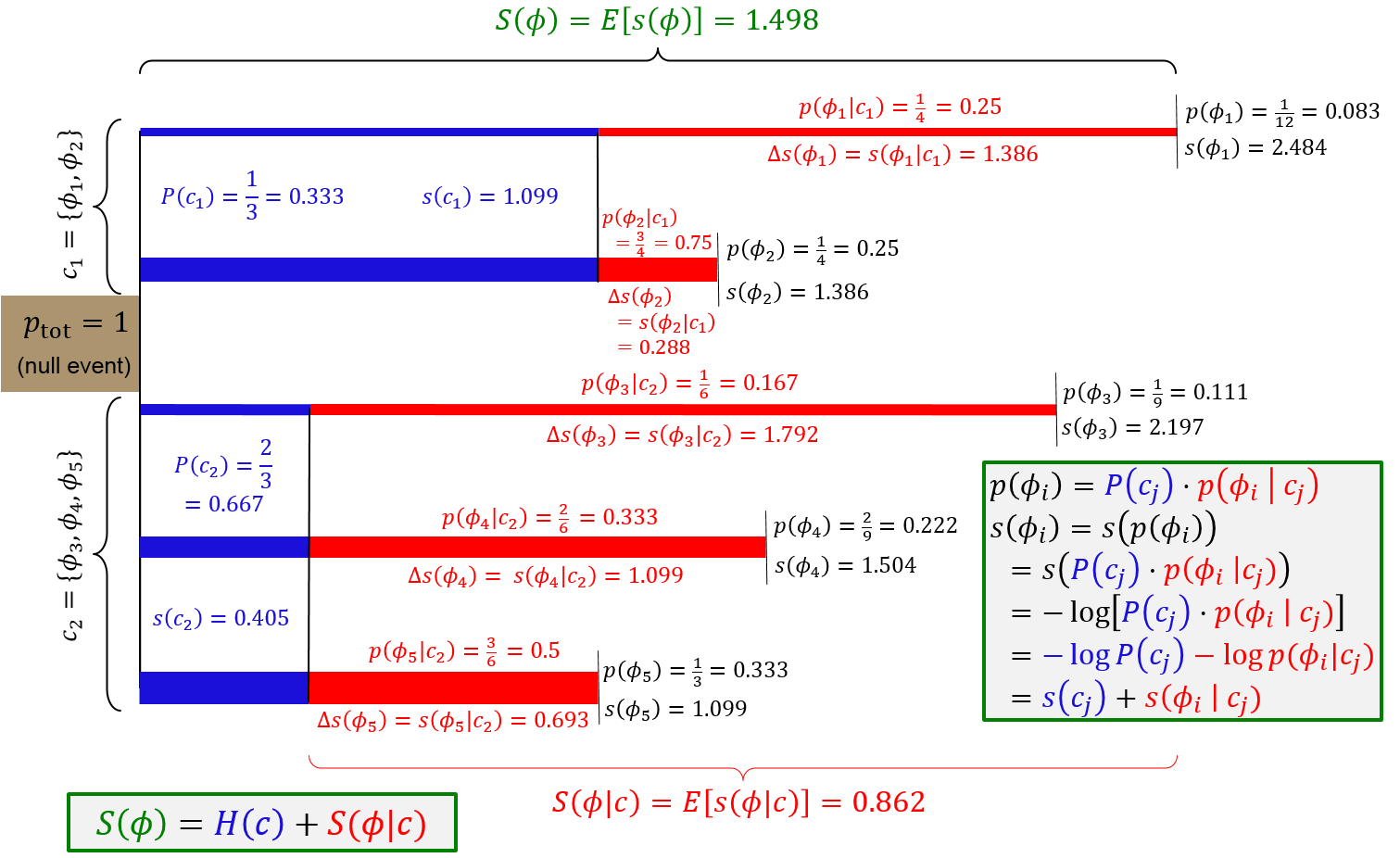}
		\caption{Detailed numerical data for the example of 
		Fig.~\ref{fig:trees}(b), during the grouping operation. In the inset, we 
		also give an analytical derivation showing that the total surprise (length 
		from root to leaf) for each branch is conserved by the grouping operation.}
		\label{fig:tree2}
	\end{figure}
	

	\begin{figure}
		\centering\includegraphics[height=3in]{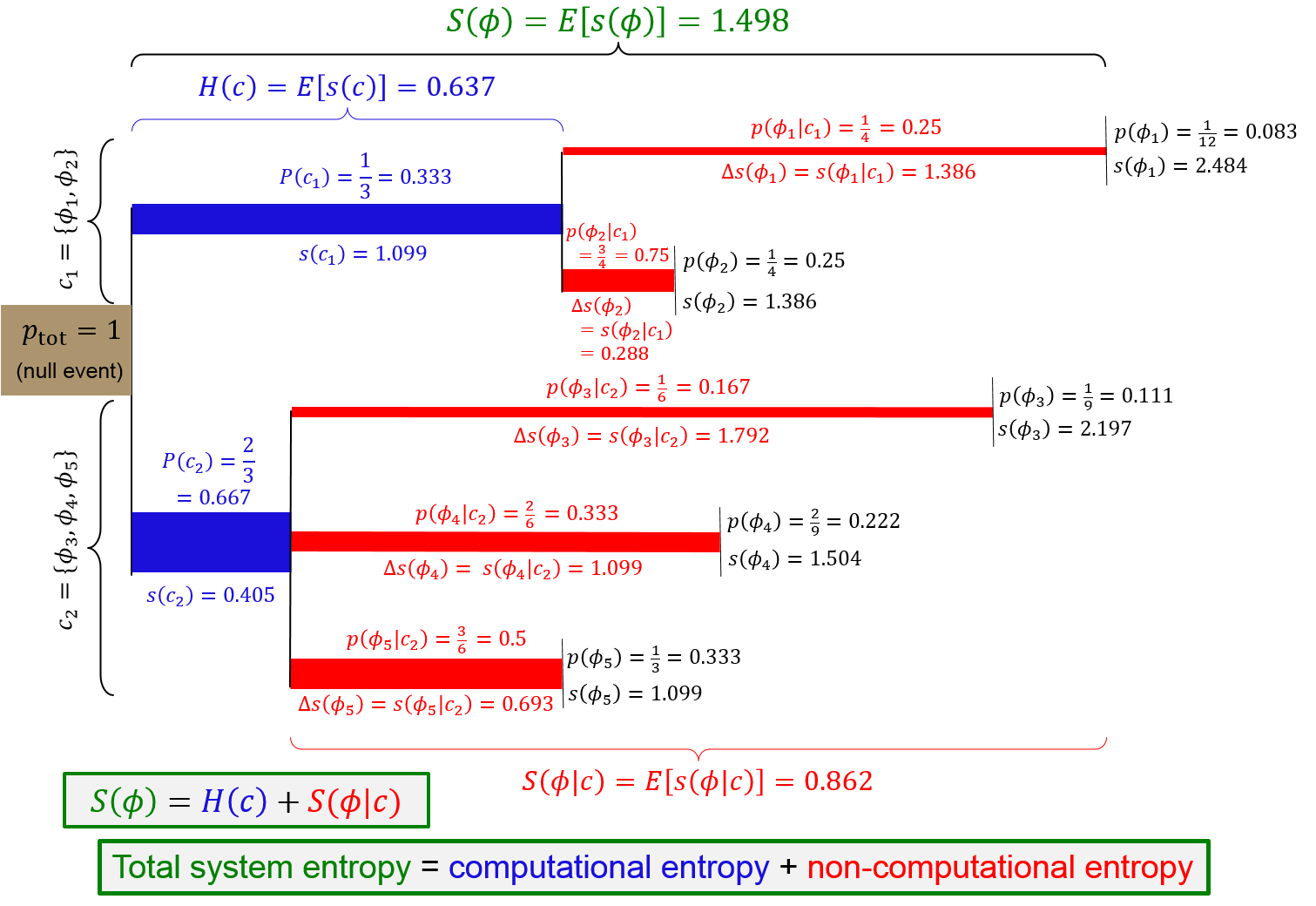}
		\caption{Detailed numerical data for the example of 
		Fig.~\ref{fig:trees}(c), resulting from the grouping operations.  The 
		Fundamental Theorem is paraphrased at bottom.}
		\label{fig:tree3}
	\end{figure}


\end{document}